\documentclass[apj]{emulateapj}
\usepackage{multirow}
\usepackage{graphics}
\usepackage{rotating}
\usepackage{enumitem}
\usepackage{amsmath}
\usepackage{natbib}
\usepackage{hyperref}
\pdfoutput=1
\hypersetup{
  hyperfootnotes=true,    
  colorlinks=true,        
  linkcolor=blue,         
  citecolor=blue,         
  urlcolor=blue,          
  breaklinks=false,        
}
\usepackage{array,booktabs,ragged2e}

\newcolumntype{R}[1]{>{\RaggedLeft\arraybackslash}p{#1}}
\newcolumntype{C}[1]{>{\centering\arraybackslash}p{#1}}

\shorttitle{On the Structure of the AGN Torus}
\shortauthors{Khim \& Yi}

\def\kms{${\rm km}~{\rm s}^{-1}$}

\newcommand{\gandalf}{{\texttt {gandalf}}}
\newcommand{\GIMtwoD}{{\texttt {GIM2D}}}

\newcommand{\specclass}{{\texttt {specClass=3}}}

\def\Halpha{\mbox{H$\alpha$\,}}
\def\Hbeta{\mbox{H$\beta$\,}}
\def\OIIILambda{[\mbox{O\,{\sc iii}}]~$\lambda 5007$}
\def\NIILambda{[\mbox{N\,{\sc ii}}]~$\lambda 6584$}
\def\MBH{\mbox{$M_{\rm BH}$}}
\def\Lbol{\mbox{$L_{\rm bol}$}}
\def\EddRatio{\mbox{$\lambda_{\rm Edd}$}}

\def\OI{[\mbox{O\,{\sc i}}]}

\def\OIII{[\mbox{O\,{\sc iii}}]}
\def\NII{[\mbox{N\,{\sc ii}}]}

\def\kms  {\hbox{${\, \rm km}~{\rm s}^{-1}$}}
\def\ergs {\hbox{${\, \rm erg}~{\rm s}^{-1}$}}

\begin{document}

\title{On the Structure of the AGN Torus through the Fraction of Optically-Selected Type 1 AGNs}

\author{Honggeun Khim and Sukyoung K. Yi}
\affil{Department of Astronomy and Yonsei University Observatory, Yonsei University, Seoul 03722, Republic of Korea: yi@yonsei.ac.kr}

\begin{abstract}
The ratio in number between unobscured (type~1) and obscured (type~2) AGNs is often used to explore the structure of the torus in the unified scheme for active galactic nuclei (AGNs). \citet{oh15} investigated the type 1 AGN fraction on two-dimensional space in terms of black hole mass ($\MBH$) and bolometric luminosity ($\Lbol$) and found that the fraction changes depending on both $\MBH$ and $\Lbol$, forming a ridge-shaped distribution. In this study, based on the up-to-date type 1 AGN catalog of \citet{oh15}, we examine how the trend of the type 1 AGN fraction in the $\MBH-\Lbol$ plane is affected by the different methods used to derive $\MBH$ and $\Lbol$, and suggest an analytic model to explain the observations. We use galaxies from the Sloan Digital Sky Survey Data Release 7 in the redshift range of $0.01 \leq z \leq 0.2$. In estimating $\Lbol$, we employ two different methods using $\OIII$ and/or $\OI$ emission lines, and find that the $\Lbol$ values obtained from the two methods agree well. We consider the $\MBH-\sigma_*$ relation, the $\MBH-L_{\rm bulge}$ relation and the single-epoch $\Halpha$-based $\MBH$ estimate in calculating $\MBH$. We find that the trends of the type 1 AGN fraction with respect to $\MBH$ and $\Lbol$ are similar for the different derivation methods for $\Lbol$ but different when using different methods to derive $\MBH$. We present a model based on the clumpy-torus scheme that reproduces the ridge-shaped distribution of the fraction parallel to the iso-Eddington ratio lines.
\end{abstract}



\section{Introduction}
\label{sec:intro}

Active galactic nuclei (AGNs) are thought to be powered by central supermassive black holes (SMBHs) and play a significant role in galaxy evolution through AGN feedback. Historically, \citet{sey43} was the first to notice a distinct class of galaxies, the so-called Seyfert galaxies which host low-luminosity AGNs, based on characteristics of optical spectra such as broad Balmer emission lines or certain asymmetric emissions. Seyfert galaxies can be classified as type 1 with broad Balmer emission lines and type 2 without broad Balmer emission lines \citep{kha74}, and further into intermediate types, Seyfert 1.2, 1.5, 1.8 and 1.9, based on the relative strengths of broad and narrow H$\beta$ emission lines \citep{ost77,ost81}.

The standard unified model of AGNs was introduced to explain their various types through orientation effects \citep{ant93,urr95}. According to the unified model, central sources of the AGNs, which consist of an accretion disk and a broad-line region (BLR) located near the SMBH, are enclosed within an optically-thick circumnuclear medium with a toroidal shape. Therefore, broad emission lines are absent in type 2 AGNs because of the obscuration of the BLR when viewed through a dusty torus, whereas they are visible in type 1 AGNs because there is no obscuration in the line of sight. Based on the structural scheme of AGNs, the unified model argues that different classes of AGNs are intrinsically the same but appear different merely due to the orientation effect.

This unified model is supported by several observations. \citet{ant85} observed scattered broad optical lines in a polarized spectrum of NGC 1068, a Seyfert 2 galaxy, which is strong evidence of a hidden BLR. Hidden BLRs have been discovered by subsequent studies for other narrow-line AGNs using polarized optical spectra \citep[e.g.,][]{mil90,tra92,tra95,mor00}. Though only about 50$\%$ of Seyfert 2 galaxies seem to possess hidden BLRs \citep[e.g.,][]{tra01,tra03,gu02}, most detection failures may be attributed to the low rate of mirror reflection or other effects that make the detection of weak polarized light difficult \citep{ale01,gu01}. In addition, broad lines observed in the infrared (IR) spectra for some narrow-line AGNs imply the existence of hidden BLRs \citep[e.g.,][]{rui94,vei99,lam17}, supporting the concept of the unified model.

It should be noted that some recent studies suggested an alternative possibility that the degree of obscuration in the central region may also be determined by large-scale clustering of AGNs \citep[e.g.,][]{all14,don14}). In our study, however, we focus on the more traditional unified model.

There have been many studies on the structure of the torus which is a key ingredient of the unified scheme. These structures can be investigated through the fraction of type 1 AGNs because the fraction is mainly determined by the geometry of the torus. In particular, the type 1 AGN fraction along bolometric luminosity, or $\OIII$ luminosity as an indicator, has been explored over various wavelength bands. For X-ray-selected AGN samples, the decreasing trend of the obscured fraction (type 2 AGN fraction) with increasing AGN luminosity was observed \citep{ued03,ste03,has04,la05,tre06,has08}.

Using IR spectra, some studies have indicated that the ratio of mid-IR to optical or bolometric luminosity declines as luminosity increases \citep{mai07,tre08,lus13}. This means that with an increase in luminosity, the covering factor of the torus decreases, and thus, the type 1 AGN fraction increases. \citet{ass13} studied the type 1 AGN fraction in mid-IR using the \textit {Wide-field Infrared Survey Explorer} \citep[WISE;][]{wri10} database, and found the same luminosity-dependent type 1 AGN fraction. Furthermore, the fraction of type 1 AGNs has explored based on radio-selected AGNs. \citet{gri04} showed an increasing quasar fraction among radio-loud galaxies (radio galaxies and quasars) with $\OIII$ luminosity. \citet{ars05} examined the opening angle of a torus based on projected sizes of radio-loud galaxies, and found that the opening angle decreases with increasing $\OIII$ luminosity, representing an increasing type 1 AGN fraction with luminosity.

The Sloan Digital Sky Survey \citep[SDSS;][]{yor00} is one of the largest optical surveys of galaxies. The SDSS database enables us to comprehensively analyze the type 1 AGN fraction of optically-selected AGNs with a large number of galaxies. \citet{sim05} constructed a sample of type 1 and type 2 AGNs based on the SDSS Data Release 2 (DR2) catalog, and found that the type 1 AGN fraction increases as a function of $\OIII$ luminosity. Using the SDSS database, \citet{oh15} developed techniques to improve the completeness of the type 1 AGN catalog by securely detecting the broad components of H$\alpha$ emission lines. They suggested an increasing trend of the type 1 AGN fraction with $\OIII$ luminosity as presented in previous studies.

In addition, by analyzing the trend of the type 1 AGN fraction in the $\MBH-\Lbol$ plane, \citet{oh15} showed that this fraction may be influenced by black hole mass ($\MBH$) as well as bolometric luminosity ($\Lbol$). They also pointed out that the fraction changes depending on the Eddington ratio, forming a ridge-shaped distribution on the plane. However, this result cannot be explained by a standard torus model, the so-called receding torus model proposed by \citet{law91} as it was devised to explain the luminosity-dependent aspect only. Thus, we need to develop a new torus model to consider effects of both $\MBH$ and $\Lbol$.

In this study, we aim to examine the effect of $\MBH$ and $\Lbol$ on the structure of the AGN torus through the fraction of type 1 AGNs, and test a simple torus model to figure out the trend of the fraction. The outline of this paper is as follows. In section~\ref{sec:sample}, we describe the sample section processes based on an up-to-date type 1 AGN catalog provided by \citet{oh15} and an emission-line diagnostic diagram for type 2 AGNs. The fraction of type 1 AGNs as a function of $\OIII$ luminosity is addressed in section~\ref{sec:fraction_OIII}. In section~\ref{sec:L_bol_M_BH}, we investigate various methods of deriving $\Lbol$ and $\MBH$. In section~\ref{sec:fraction_2D}, we show the type 1 AGN fraction on the $M_{\rm BH}-L_{\rm bol}$ plane for our sample, and examine how trends of the fraction can differ depending on the methods of deriving $\MBH$ and $\Lbol$. Based on the clumpy torus model \citep{nen02,nen08a,nen08b}, we suggest an analytic model of the AGN torus to explain the observed fraction of type 1 AGNs (section~\ref{sec:discussion}). Our results are summarized in section~\ref{sec:summary}. Throughout this paper, we assume the following cosmological parameters: H$_0$ = 70 \kms~Mpc$^{-1}$, $\Omega_{\rm M}$ = 0.3 and $\Omega_\Lambda$ = 0.7.


\section{Sample Selection}
\label{sec:sample}
 
To study the fraction of type 1 AGNs, we utilize the SDSS Data Release 7 \citep[DR7;][]{aba09}. The SDSS observation was performed using a wide-field 2.5~m telescope located at Apache Point, New Mexico, the USA \citep{gun06}, covering one-quarter of the sky. We use the ``main'' galaxy sample for both type 1 and type 2 AGNs, instead of the ``quasar'' sample (see section~\ref{sec:type1}) because the main galaxy sample is supposed to be spectroscopically more complete for galaxies brighter than 17.77 in extinction-corrected $r$-band Petrosian magnitude \citep{str02}. For spectroscopic information, we use the OSSY catalog \citep{oh11}, which aims to provide improved spectroscopic measurements for an individual galaxy at $z \leq 0.2$ based on the SDSS DR7 single-aperture spectrum. Matching the main galaxy sample with the OSSY catalog, we choose galaxies in the redshift range of $0.01 \leq z \leq 0.2$. We do not include galaxies at $z < 0.01$ to avoid saturated spectra. As a result, we have a population size of 637,511 galaxies.

\subsection{AGN classification}
\label{sec:classification}


\subsubsection{Type 1 AGNs}
\label{sec:type1}
Quasars identified by SDSS pipelines, including both type 1 quasars and less luminous type 1 AGNs, are categorized via photometric and subsequent spectroscopic selection processes. Unlike stars showing roughly blackbody spectra, quasars demonstrate power-law continua with strong emission lines. The ``photometric'' quasar candidates are selected, therefore, based on different locations of stars and quasars on color-color diagrams with point-spread function (PSF) magnitudes. After follow-up spectroscopic observations, the objects displaying a well-defined broad emission line with a full width at half maximum (FWHM) greater than $1,000$ $\kms$ are designated as quasars \citep{ade07} with a flag $\specclass$ for low-redshift quasars \citep{sto02}. For the SDSS DR7, there are 4,125 quasars with $\specclass$ at $z<0.2$.

\citet{oh15} created a new catalog of type 1 AGNs based on all spectroscopic objects including both the main galaxy and quasar samples of the SDSS. Target galaxies were drawn from the SDSS DR7 at $z<0.2$. The catalog was constructed through several steps. First, they identified type 1 AGN candidates from the target sample by analyzing a flux ratio at the near and far shoulders of the H$\alpha$ emission line which could be large under the existence of the broad component. For the type 1 AGN candidates, FWHM(broad H$\alpha$) $>800$ $\kms$, and $A/N$(broad H$\alpha$) $>3$, where $A$ is the best-fitting amplitude of the emission line and $N$ is the residual noise, were set as the selection criteria of potential type 1 AGNs. Here, the amplitude and noise were measured using the $\gandalf$ fit which is a galaxy spectral line fitting algorithm \citep{sar06} considering stellar kinematics measurements \citep[pPXF;][]{cap04}. Finally, they introduced an ``areal ratio'' criterion to exclude doubtful type 1 AGNs among those with weak broad-line features. To rule out the possible misclassification, they computed the extent of which the broad feature on the right-hand side exceeds the wing of the Gaussian profile of \NIILambda \, emission compared to the noise level. With these robust sampling processes, the catalog presents 5,553 type 1 AGNs, recovering $\sim$90\% of the $\specclass$ objects with the same conditions used in \citet{oh15} and newly adding 1,846 type 1 AGNs. The detailed selection procedures are described in \citet{oh15}.

In this study, we use the catalog provided by \citet{oh15} as the basis of our type 1 AGN sample. As mentioned above, we only select galaxies in the main galaxy sample (i.e., $r_{\rm petro} < 17.77$) in order to achieve a high level of spectroscopic completeness. This reduces the sample size by 23\%, and gives us 4,268 type 1 AGNs.


\begin{figure}
\includegraphics[width=8cm, angle=270, trim=0 0.5cm 0 0, clip]{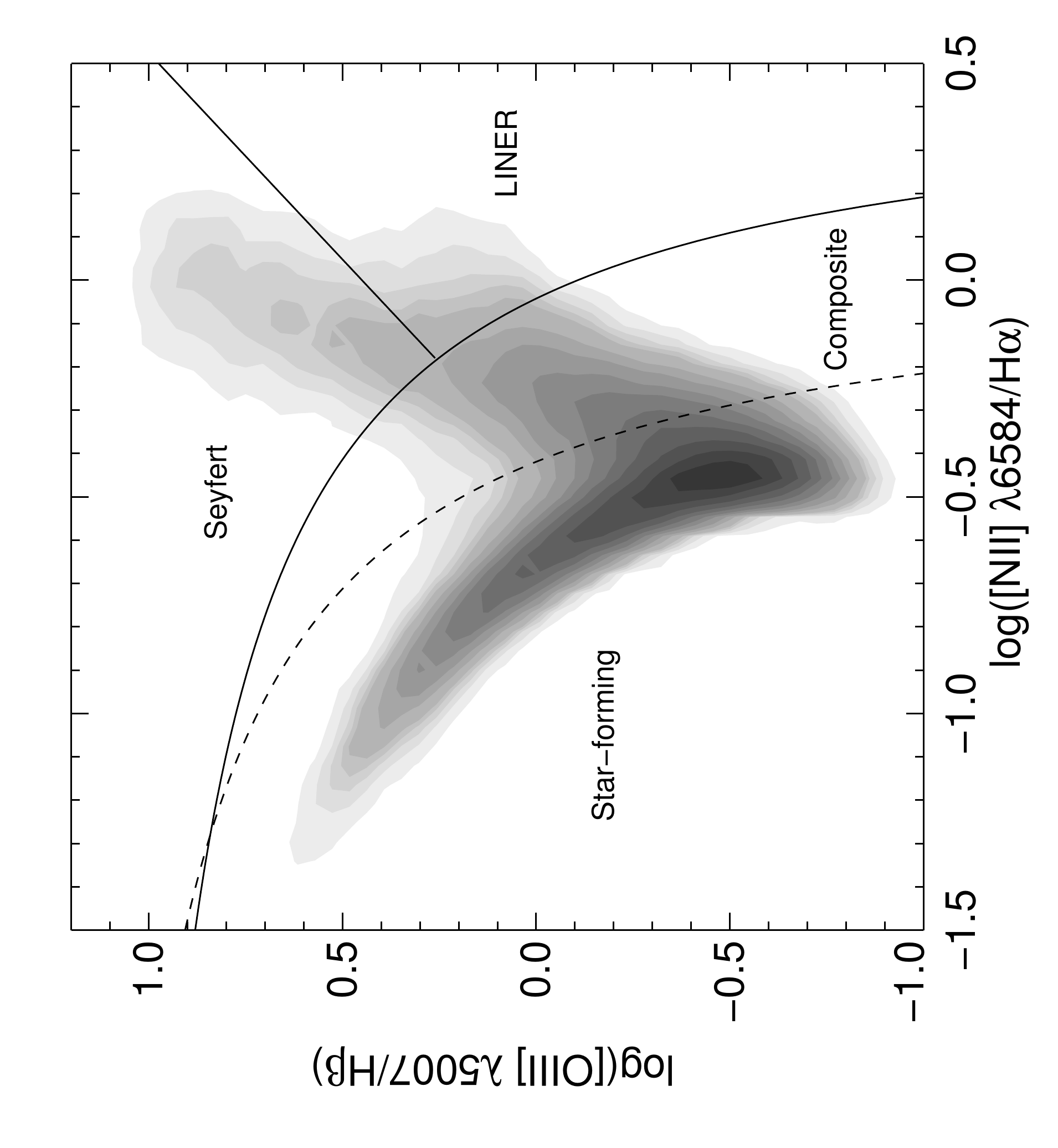}
\caption[BPT diagram]{BPT diagram with \Halpha, \Hbeta,  \NIILambda \ and \OIIILambda \ emission lines. The diagram shows galaxies with $A/N>3$. The darker region indicates a high density of galaxies. The dashed curve indicates the demarcation line from \citet{kew01}, the solid curve is from \citet{kau03} and the solid straight line is from \citet{sch07}. The corresponding regions divided by the demarcation lines are noted.}
\label{fig:bpt}
\end{figure}

\subsubsection{Type 2 AGNs}
\label{sec:type2}
Unlike type 1 AGNs which show broad emission lines, type 2 AGNs do not show conspicuous features against star-forming galaxies. Therefore, type 2 AGNs are generally identified by analyzing the relative strengths of the narrow emissions. In this study, type 2 AGNs are selected from the main galaxy sample using the BPT diagram of  \NIILambda/H$\alpha$ and \OIIILambda/H$\beta$ flux ratios devised by \citet{bal81}. We use three demarcation lines to classify the emission-line galaxies into star-forming, composite, LINER (Low Ionization Nuclear Emission-line Region), and Seyfert galaxies \citep{vei87,kew01,kau03,sch07} as shown in Fig.~\ref{fig:bpt}. The fluxes of the emission lines are taken from the OSSY catalog and extinction-corrected. The $A/N$ criteria ($A/N>3$) for the four emission lines, \Halpha, \Hbeta,  \NIILambda \, and \OIIILambda, are applied, and 207,164 galaxies are chosen from the OSSY catalog. Fig.~\ref{fig:bpt} shows the galaxies with $A/N>3$ for all four emission lines, of which 73.8\% are star-forming, 17.4\% are composite, 2.7\% are LINER and 6.1\% are Seyfert galaxies.

\begin{figure}
\centerline{\includegraphics[width=8cm, angle=270, trim=0 0.5cm 0 0, clip]{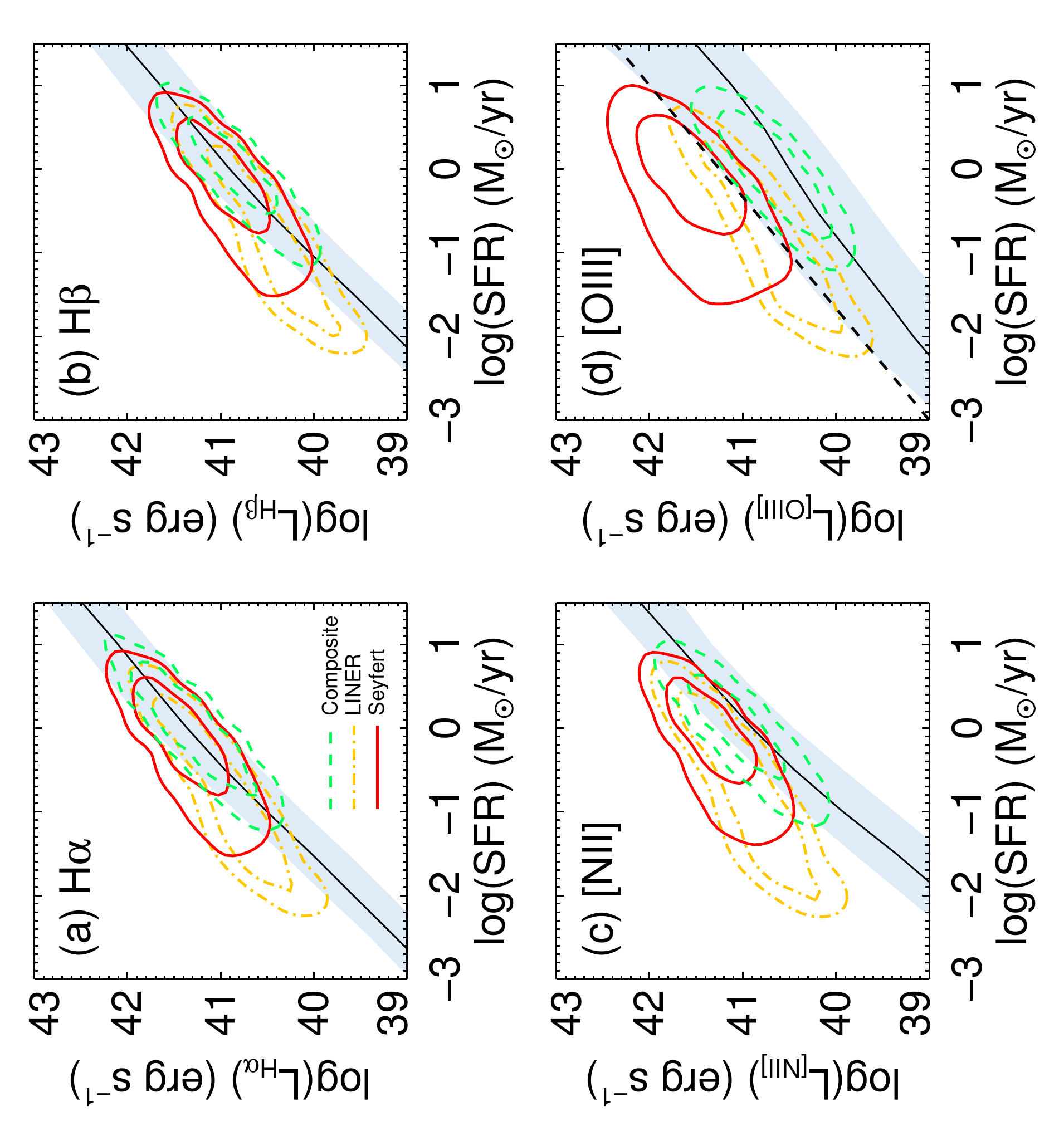}}
\caption[Luminosity of emissions for the SFR]{Luminosities of four emission lines as a function of the SFR. Panels (a), (b), (c) and (d) show the H$\alpha$, H$\beta$, \NII \ and \OIII \ emission line, respectively. The x-axis is the SFR and the y-axis is the luminosity of the emission. In each panel. the relation between the SFR and the luminosity of the star-forming galaxies is shown as a solid black line (median) and a shaded region (2$\sigma$). Red, orange and green contours (0.5$\sigma$ and 1$\sigma$) indicate Seyfert, LINER and composite galaxies, respectively. The dashed line in panel (d) is the demarcation line we use to choose AGNs from the LINER and the composite region.}
\label{fig:sfr_emission}
\end{figure}

Before we identify type 2 AGNs, we exclude the type 1 AGNs defined in section~\ref{sec:type1}. The narrow emission lines of Seyfert galaxies are considered to be mainly from AGNs, whereas in the case of composite galaxies, the emissions from star formation may be comparable to those from AGNs. Hence, when constructing the type 2 AGN sample, identifying AGNs in the composite region is challenging. To pick out AGNs from the composite region, we calculate the mean strengths of the four emission lines (\Halpha, \Hbeta,  \NIILambda \ and \OIIILambda) of star-forming galaxies as a function of the star-formation rate (SFR) and compare them with those of composite galaxies. Assuming that the emission-line strengths of the star-forming galaxies are purely from the star formation, we can define composite galaxies as ``AGNs'' when the galaxies show stronger emissions than star-forming galaxies with the same SFRs. In this approach, we utilize the SFR within the SDSS fiber from the MPA-JHU DR7 catalog, which is estimated following the method used in \citet{bri04}.

Fig.~\ref{fig:sfr_emission} shows the de-reddened luminosities of emission lines as a function of the SFR. Panels (a), (b), (c) and (d) are the H$\alpha$, H$\beta$, \NII \ and \OIII \ emission lines, respectively. Solid black lines indicate the median relations between the SFR and luminosity in star-forming galaxies and the shaded regions show the 2$\sigma$ ranges. All panels demonstrate that the strengths of emission lines of star-forming galaxies monotonically increase with SFR. Green, orange and red contours in each panel correspond to composite, LINER and Seyfert galaxies with 0.5$\sigma$ and 1$\sigma$, respectively. For all panels, composite galaxies (green contours) are aligned with the relation between the SFR and emission lines of the star-forming galaxies, indicating that star formation is the dominant mechanism for most composite galaxies. In the case of Balmer lines (panels (a) and (b)), even Seyfert galaxies overlap with the shaded zones, meaning that the Balmer emissions from AGNs are not much stronger than those from star formation. On the other hand, for the forbidden lines (panels (c) and (d)), Seyfert galaxies show a clear offset from the shaded regions, especially for $\OIII$ luminosity. We find that 88.1\% of the Seyfert galaxies are at the upper side of the shaded area in panel (d), and hence, the boundary of the area (dashed line in panel (d)) is a good criterion for discriminating AGNs in the composite region. This criterion is given by equation~(\ref{eqn:demarcation}).
\begin{align}
\log \bigg(\frac{L_{[\mbox{O\,{\sc iii}}]}}{\ergs}\bigg) > 0.75\log \bigg(\frac{SFR}{M_\odot \; yr^{-1}}\bigg)+41.25.
\label{eqn:demarcation}
\end{align}

In addition, there is discussion as to whether the features of LINERs originate from AGNs \citep{hec80,fer83,ho93}, star formation \citep{ter85,shi92} or post-AGB stars and corresponding planetary nebulae \citep{di90,bin94,sar10,sin13}. In panels (a) and (b), LINERs show similar Balmer-emission luminosities to those of star-forming galaxies at high SFR. Panel (d) also illustrates that around half of the LINERs have SFRs comparable to those of star-forming galaxies (shaded zone), meaning that there could be some LINERs for which star formation is significant. Therefore, we take into account the criterion in panel (d) to securely select AGNs from the LINERs.

By applying the above criterion (equation~(\ref{eqn:demarcation})) to LINERs and composite galaxies, 53.9\% of LINERs (2,724) and 9.2\% of composite galaxies (3,028) are chosen as type 2 AGNs. After adding the type 2 Seyfert galaxies (9,406) to the AGN sample, we have 15,158 type 2 AGNs. For our sample, the ratio between the number of type 1 and type 2 AGNs is $\sim$1:4 which is consistent with the results of the previous studies \citep[e.g.,][]{mai95}.


\section{Type 1 AGN Fraction as a Function of $\OIII$ Luminosity}
\label{sec:fraction_OIII}

\begin{figure}[t]
\centerline{\includegraphics[width=8cm, angle=270, trim=0 0.5cm 0 0, clip]{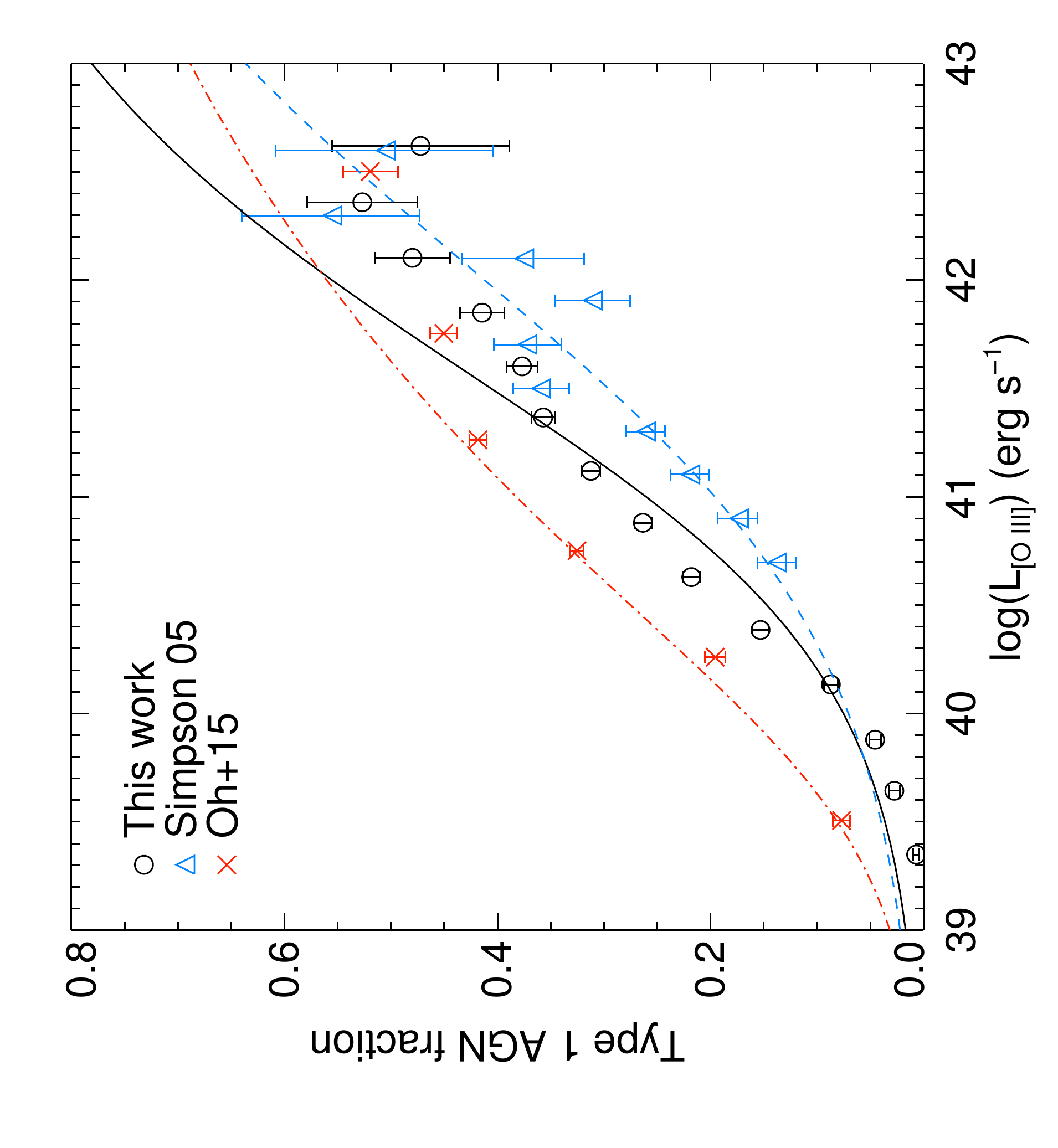}}
\caption[Type 1 AGN fraction as a function of the observed $\OIII$ luminosity]{Type 1 AGN fraction as a function of the observed $\OIII$ luminosity without reddening correction. The x-axis represents $\OIII$ luminosity and the y-axis represents the type 1 AGN fraction. The fractions of this study, \citet{sim05} and \citet{oh15} are shown by black circles, blue triangles and red crosses, respectively. Each line indicates the fitting line for each data sample with the same color.}
\label{fig:frac_OIII}
\end{figure}

In this section, we discuss the variation in the type 1 AGN fraction\footnote{The fraction of type 1 AGNs is defined as follows: $f_{1} = \frac{N_{1}}{N_{1}+N_{2}}$, where $N_{1}$ and $N_{2}$ are the number of type 1 and type 2 AGNs, respectively.} with $\OIII$ luminosity. Fig.~\ref{fig:frac_OIII} shows the type 1 AGN fraction as a function of $\OIII$ luminosity for our sample (black circles). The results of previous studies based on the SDSS database are represented by blue triangles \citep{sim05} and red crosses \citep{oh15}. We use $\OIII$ luminosity without reddening correction to directly compare the results. It is clear that the type 1 AGN fraction increases with $\OIII$ luminosity for all cases, but there are differences in the fraction between samples. \citet{oh15} present an increasing trend of the type 1 AGN fraction from 0.1 to 0.5 as $\OIII$ luminosity increases from $10^{39.5}$ to $10^{42.5}$ \ergs, showing greater fraction values than this study. This is because we select more type 2 AGNs in the LINER and the composite region, and apply the magnitude cut $r_{\rm petro} < 17.77$ (which is not used by \citealt{oh15}), resulting in a relatively small number of type 1 AGNs in our sample. The overall fraction of \citet{sim05} is slightly lower than that of this study, which shows the type 1 AGN fraction increasing from 0.15 to 0.5 with $\OIII$ luminosity increasing from $10^{40.5}$ to $10^{42.5}$ \ergs. He used objects with $r_{\rm petro} < 17.77$ as we do, and the difference in the fraction may have originated from a different way of defining type 1 AGNs. \citet{oh15}, which is the basis of our sample, only considered the presence of H$\alpha$ broad emission lines when choosing type 1 AGNs, whereas \citet{sim05} defined objects as type 1 AGNs when there were clear broad components in both H$\alpha$ and H$\beta$ emissions. In other words, the strict criterion in \citet{sim05} may have decreased the number of type 1 AGNs.

It is important to understand the impact of selection criteria, such as redshift, $A/N$ or FWHMs of broad emission lines, on the fraction determination. Ffor the impact of redshift as a selection criterion, the type 1 AGN fractions are reasonably stable within Poisson errors. For high-luminosity bins, the impact is larger but is roughly in the level that can be explained as the effect of small number statistics. Second for the criteria regarding broad lines, we find the fractions become lower for stricter FWHM cuts (e.g., FWHM(broad H$\alpha$) $>2,000$ \kms) or stricter $A/N$ cuts (e.g., $A/N$(broad H$\alpha$) $>5$) as expected; but the effects are still marginal and the results are in reasonable agreement within Poisson errors.

To explain the luminosity-dependent type 1 AGN fraction, \citet{law91} proposed the receding torus model. Dust grains in the torus can exist only when the optical and UV radiation from the SMBH are not strong enough to destroy it by evaporation. Thus, the inner radius of the torus is determined by the radius where the temperature of the dust, decreasing as a function of the distance from the central black hole, is equal to the evaporation temperature. As a result, the inner radius is further pushed away for luminous AGNs. The receding inner radius of the torus is calculated by theoretical models, suggesting $R_{\rm in} \propto L_{\rm AGN}^{0.5}$ \citep{bar87,nen02}. Because the fraction of type 1 AGNs is mainly determined by the geometry of the torus, \citet{law91} constructed the receding torus model to explain the higher type 1 AGN fraction for luminous AGNs by assuming the luminosity-dependent inner radius and constant height of the torus. This simple model succeeds in reproducing the trend of the type 1 AGN fraction in \citet{wil00} and \citet{gri04} based on radio-selected AGNs.

Though this standard model is successful for some studies, \citet{sim05} found that the standard receding torus model with the constant height of the torus is insufficient to explain the type 1 AGN fraction based on the SDSS database. Allowing for variation in height with luminosity ($h \propto L^\xi$), he suggested a modified receding torus model, $f_{1} = 1-(1+3(L_{\scriptsize \OIII}/L_0)^{1-2\xi})^{-0.5}$ where $L_{\scriptsize \OIII}$ is $\OIII$ luminosity and $L_0$ is the luminosity where the type 1 AGN fraction is 0.5. He found that this revised model presented a better fit (blue dashed line) with $\xi=0.23$ ($\chi^2_\nu=1.51$) than the standard model. Through a careful selection of the FR IIs and quasars in \citet{gri04}, \citet{ars05} also concluded that a torus model with a luminosity-dependent height of the torus is required, suggesting $h \propto L^{0.26}$ which is in reasonable agreement with the result of \citet{sim05}.

The modified model, however, is not applicable to the AGN sample of \citet{oh15}. To determine the trend of the type 1 AGN fraction for their sample, \citet{oh15} introduced a new parameter to increase the flexibility of the modified receding torus model. The model with the additional parameter fits the trend of the fraction better than the modified model of \citet{sim05}, but the fit is still poor for luminous AGNs ($L_{\scriptsize \OIII} \gtrsim 10^{41.5}\ergs$), giving $\chi^2_\nu=8.3$ as shown by the red dot-dashed line in Fig.~\ref{fig:frac_OIII}.

We perform a $\chi^2$ fit for our sample with the modified receding torus model suggested by \citet{sim05}. The result of the fit is shown by black solid line in Fig.~\ref{fig:frac_OIII}. The best-fit values of $\xi$ and $L_0$ are 0.16 and $10^{41.81}\ergs$ with $\chi^2_\nu=16.8$. It is clear that the fitting line does not follow the trend of the type 1 AGN fraction and the $\chi^2_\nu$ value being far from unity indicates that the model does not explain observations well. This result suggests that the luminosity-dependent model may not be sufficient and the consideration of the effects of both $\MBH$ and $\Lbol$ on the fraction of type 1 AGNs may be essential as proposed by \citet{oh15}. In the following sections, we investigate various methods of deriving $\Lbol$ and $\MBH$, and analyze the type 1 AGN fraction in the $M_{\rm BH}-L_{\rm bol}$ plane.


\section{Derivation of $L_{\rm bol}$ and $M_{\rm BH}$}
\label{sec:L_bol_M_BH}


\subsection{Derivation of $L_{\rm bol}$}
\label{sec:L_bol}
As a tracer of bolometric luminosity, specific ranges of continua or certain emission lines can be used. For example, X-ray emitted from the corona near the accretion disk can be used as an indicator of $\Lbol$ \citep[e.g.,][]{mar04}. Because emission-line strengths in the visible spectrum are available for our full sample, we use them to estimate $\Lbol$ despite some uncertainties the method may bear \citep{ber15}. Relatively strong emission lines such as the $\OIII$ emission are preferred when estimating $\Lbol$ \citep[e.g.,][]{hec04,lam09,net09}. We consider the following two relations for calculating the bolometric luminosity:
\begin{align}
L_{\rm bol} &= 3500 \times L_{\rm{[O\,{\textsc {iii}}] \ \lambda5007}}^{\rm uncor},
\label{eqn:L_bol_uncor}
\end{align}
\begin{align}
\log L_{\rm bol} &= 3.80+0.25\log L_{\rm{[O\,{\textsc {iii}}] \ \lambda5007}}^{\rm cor} \notag \\ & \quad +0.75\log L_{\rm{[O\,{\textsc {i}}] \  \lambda6300}}^{\rm cor},
\label{eqn:L_bol_OI}
\end{align}
\noindent where $L_{\rm{[O\,{\textsc {iii}}] \ \lambda5007}}^{\rm uncor}$ is the $\OIII$ luminosity without reddening correction and $L_{\rm{[O\,{\textsc {iii}}] \ \lambda5007}}^{\rm cor}$ and $L_{\rm{[O\,{\textsc {i}}] \  \lambda6300}}^{\rm cor}$ are the reddening-corrected luminosities for the $\OIII$ and $\OI$ emission lines in $\ergs$. Equation~(\ref{eqn:L_bol_uncor}) is from \citet{hec04} and equation~(\ref{eqn:L_bol_OI}) is from \citet{net09}. The uncertainties of bolometric luminosity are at least 0.4 for the two relations. Here, reddening correction was performed using the Balmer decrement by \citet{oh11}. \citet{hec04} used the reddening-uncorrected $\OIII$ luminosity to avoid uncertainties in correcting extinction. \citet{net09} found that the method using the $\OIII$ emission line can underestimate $\Lbol$ of LINER-class AGNs, while the method using combination of the $\OI$ and $\OIII$ emission lines can estimate $\Lbol$ more properly for both high- and low-ionization AGNs.

\begin{figure}[t]
\centerline{\includegraphics[width=8.5cm, angle=270]{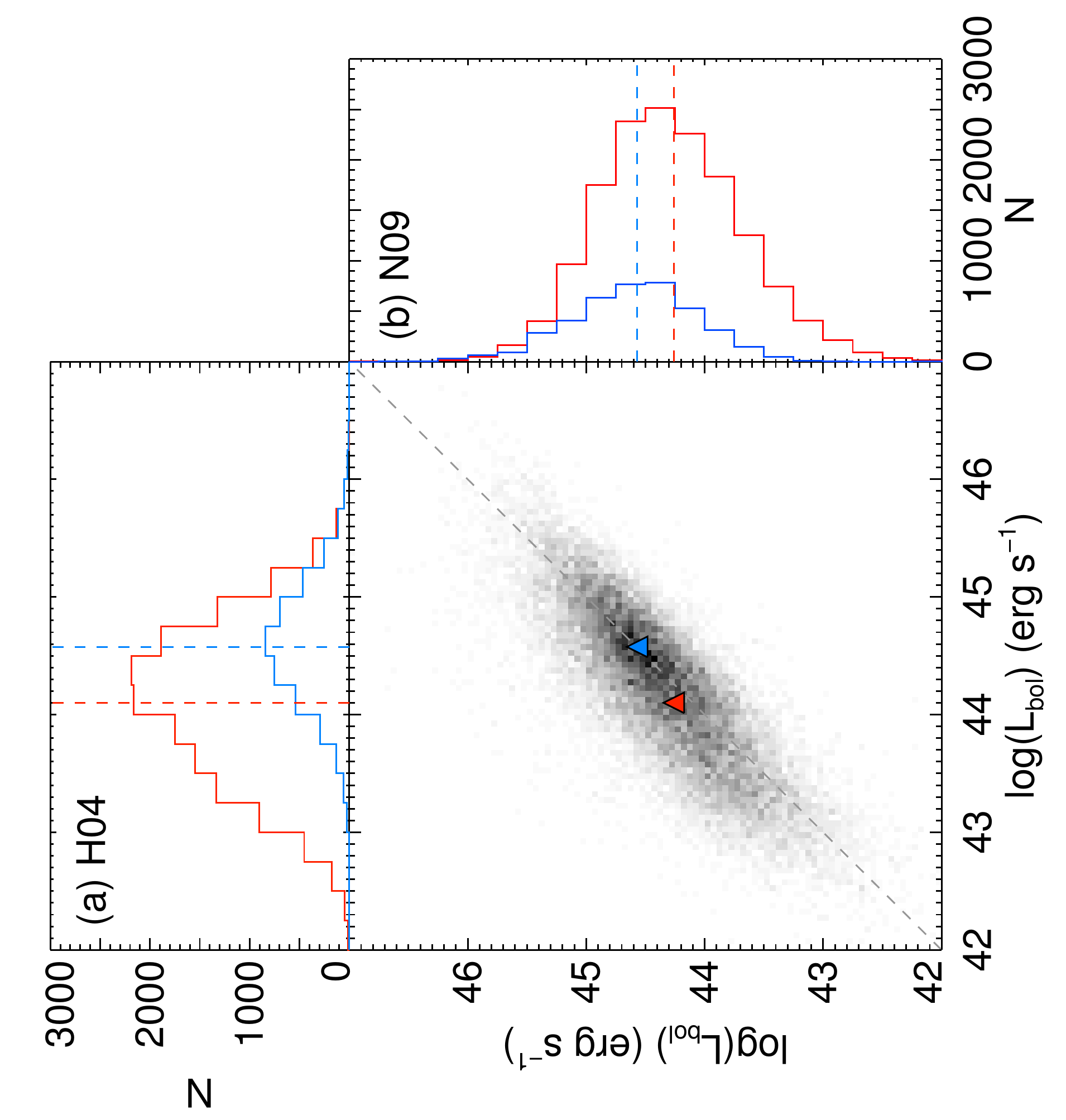}}
\caption[Distributions and relations of the $L_{\rm bol}$]{Distributions of the bolometric luminosities derived from equations~(\ref{eqn:L_bol_uncor}) (panel (a)) and (\ref{eqn:L_bol_OI}) (panel (b)) and their comparison in grayscale. Type 1 and type 2 AGNs are represented in blue and red, respectively. The mean bolometric luminosity for each type is indicated by dashed lines for the histogram diagrams and triangles for the grayscale diagram. One-to-one lines are presented by the gray dashed line.}
\label{fig:L_bol_compare}
\end{figure}

We compare the $\Lbol$ values derived by the two equations. Fig.~\ref{fig:L_bol_compare} shows the distributions of $\Lbol$ and relations between the distributions. Panels (a) and (b) correspond to the distributions of $\Lbol$ derived using equations~(\ref{eqn:L_bol_uncor}) and~(\ref{eqn:L_bol_OI}), respectively. In each panel, blue and red colors indicate type 1 and type 2 AGNs, respectively. In both panels, the distributions show an offset between the two types, with higher fractions of type 1 AGNs being expected at high-luminosity bins. The mean luminosity presented by the dashed lines also shows that the mean luminosity of type 1 AGNs is brighter than that of type 2 AGNs.

The mean luminosities of each AGN type derived according to different methods are consistent, with a value of around $10^{42.5}\ergs$ for type 1 AGNs and around $10^{42.2}\ergs$ for type 2 AGNs. Comparisons of the luminosity from each method are represented in grayscale, with mean values indicated by triangles. The gray-level density plot shows that the relations between different methods are tight and lie on the one-to-one line (black dashed lines). Small scatters may be ascribed to the uncertainty of $\sim$0.4 when deriving $\Lbol$ using the scaling relations. A slight offset from the one-to-one line is observed in the low-luminosity bins, showing that $\Lbol$ estimated from $\OIII$ luminosity is slightly smaller than that estimated from the combination of $\OI$ and $\OIII$ luminosities. Because about 20\% of our sample consist of LINERs which are generally less luminous in $\OIII$ luminosity than Seyfert galaxies, the small discrepancy may be because $\Lbol$ derived from $\OIII$ luminosity is underestimated for low-ionization AGNs as \citet{net09} pointed out. However, the difference is not significant and the bolometric luminosities obtained by the two methods are consistent within the uncertainties of the scaling relations. Thus, the tight relations between the two methods indicate that the analysis of the type 1 AGN fraction may be less affected by the choice of the method to determine $\Lbol$.


\subsection{Derivation of $M_{\rm BH}$}
\label{sec:M_BH}


\begin{figure}
\centerline{\includegraphics[width=8.5cm, trim=0 0.2cm 0 0, clip]{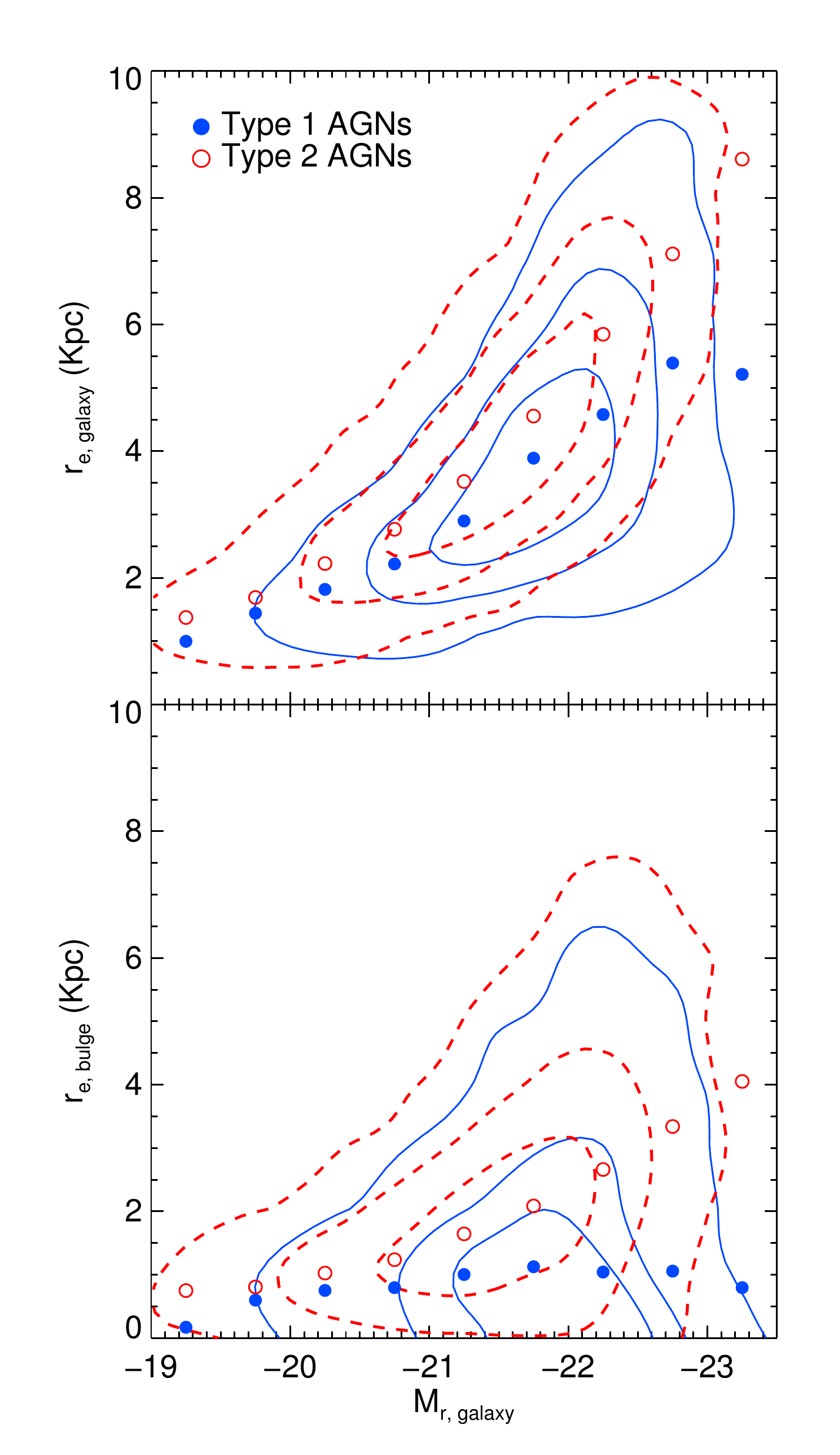}}
\caption[Effective radii against the galaxy magnitude for type 1 and type 2 AGNs]{Effective radii of galaxies (top) and bulges (bottom) against the galaxy magnitude for type 1 (blue) and type 2 (red) AGNs. The contours for type 1 (solid) and type 2 AGNs (dashed) correspond to 0.5$\sigma$, 1$\sigma$ and 2$\sigma$ from inside out, and filled circles represent the median effective radius in each galaxy-magnitude bin.}
\label{fig:eff_radius}
\end{figure}

\label{sec:various_relations}
Since the correlation between the SMBH mass and the bulge luminosity was first noticed by \citet{kor95}, many studies have investigated the relation between black hole mass and bulge properties. The velocity dispersion \citep[e.g.,][]{fer00,tre02,gul09,kor13,mcc13}, stellar and dynamic mass \citep[e.g.,][]{mag98,mar03,har04,ben11}, luminosity \citep[e.g.,][]{kor95,mcl02,ben09,kor13}, light concentration \citep[e.g.,][]{gra01} and S\'{e}rsic index \citep[e.g.,][]{grad07,bei12,vik12,sav13} have been explored to examine their relations with black hole mass. The scaling relations, however, are not consistent. For example, the slopes of the scaling relations for bulge luminosity are different ranging from -$0.5$ to -$0.3$ \citep[e.g.,][]{mcl02,mar03,gra07,ben09,gul09,san11,vik12,kor13,mcc13} depending on the sample and the method used to derive the scaling relation. As a result, black hole masses estimated from different scaling relations are inconsistent even for the same bulge luminosity. In this study, we inspect the impact of using different scaling relations.

Several studies have performed photometric decompositions for SDSS images, enabling us to utilize the structural parameters of bulges of SDSS galaxies. \citet{sim11} conducted two-dimensional bulge+disk decompositions in the $g$ and $r$-bands using the $\GIMtwoD$ software package \citep{sim02} for about a million galaxies including the main galaxy sample. They present effective radii, S\'{e}rsic indices and magnitudes of bulge and disk components of galaxies in the SDSS DR7. Other studies of the structural parameters of the bulges of SDSS galaxies have covered a narrow redshift range \citep[e.g.,][]{lac12,kim16} or did not contain $\specclass$ objects \citep[e.g.,][]{men14,mee15}. Hence, we use the catalog of \citet{sim11} to investigate the photometric properties of bulges for type 1 and type 2 AGNs.

The effective radius, $r_{\rm e}$, is a key factor in determining the black hole mass using the scaling relations with the dynamical bulge mass given by $M_{\rm dyn,bulge} \propto r_{\rm e}\sigma_{\rm e}^2/G$ \citep[e.g.,][]{mar03,cap06,wol10}. Fig.~\ref{fig:eff_radius} shows the effective radii of galaxies (top panel) and bulges (bottom panel) against the $r$-band absolute magnitude of galaxies for type 1 and type 2 AGNs, which are colored in blue and red, respectively. Filled circles indicate the median effective radii for each magnitude bin, and contours correspond to 0.5$\sigma$, 1$\sigma$ and 2$\sigma$. The effective radius of the galaxy is calculated with the Petrosian radius containing 50\% of the flux and the concentration index following the scheme described in \citet{gra05}, and the bulge effective radius is from \citet{sim11}. Here, we convert the bulge effective radius along the major axis into the equivalent circular effective radius \citep{ber03}.

The top panel in Fig.~\ref{fig:eff_radius} shows that the effective radii of galaxies indicated by contours increase for bright galaxies as expected by the galaxy luminosity-size relation \citep[e.g.,][]{she03,tru04}, though the contours for type 1 and type 2 AGNs demonstrate systematic offsets. The discrepancies between the two types are also shown by filled circles and become significant as the luminosity increases. In the bottom panel, contours show that the effective radius of the bulge for type 2 AGNs is systemically larger than that of type 1 AGNs. The median effective radius for type 2 AGNs (filled red circles) also increases with galaxy luminosity. Filled blue circles, however, are almost constant regardless of the galaxy magnitude. This may not be realistic and implies the difficulty of measuring the effective radius of type 1 AGNs accurately due to the AGN contributions from the central regions. Due to this uncertainty, the scaling relations for the dynamical mass are not included in our analysis.


\begin{figure}[t]
\centerline{\includegraphics[width=8.2cm, angle=270, trim=0 0.5cm 0 0, clip]{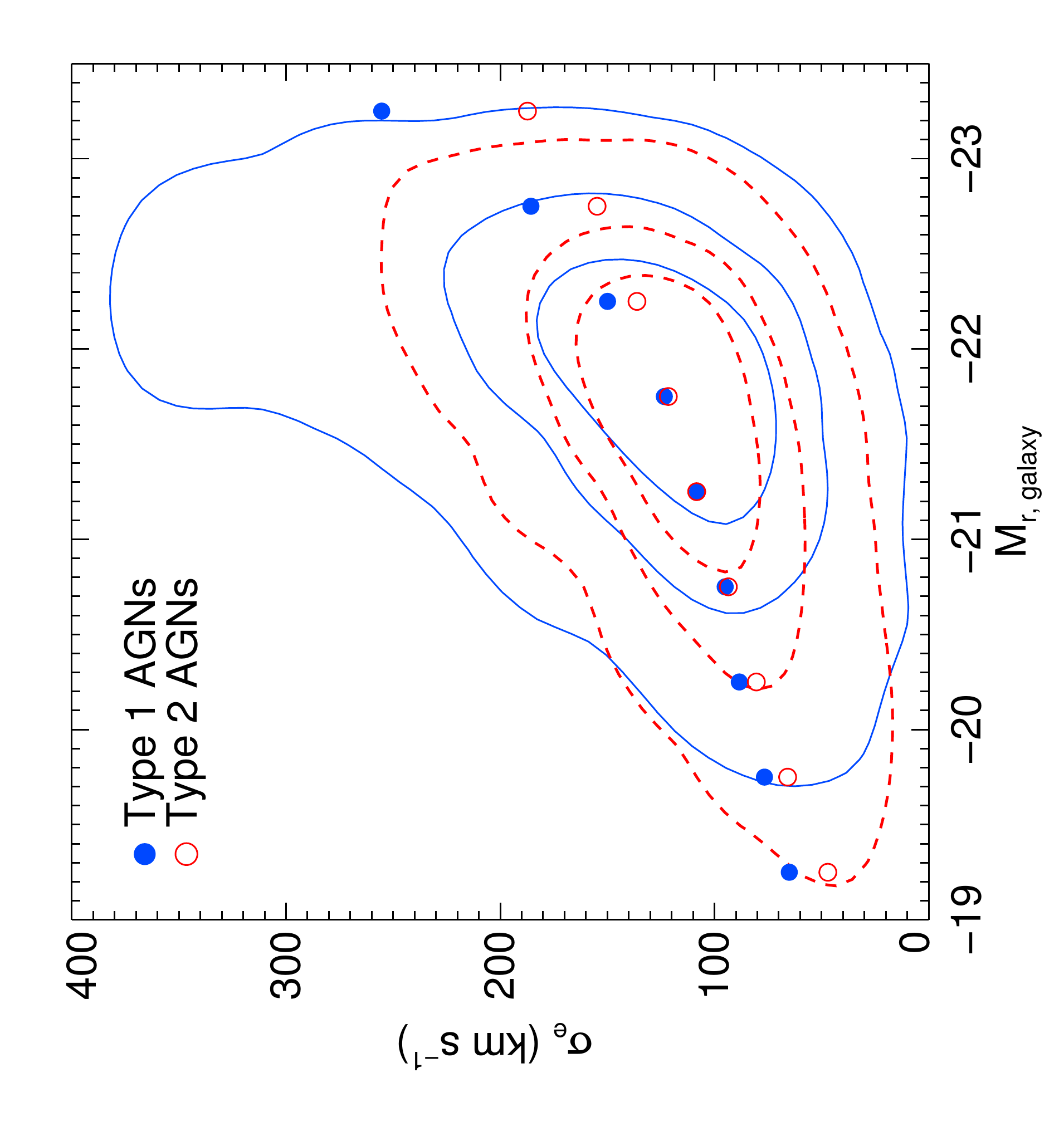}}
\caption[The distribution of the black hole mass derived from different $M_{\rm BH}-\sigma_*$ relations]{Velocity dispersion versus galaxy magnitude for type 1 and type 2 AGNs. The format is the same as that in Fig~\ref{fig:eff_radius}.}
\label{fig:velocity_disp}
\end{figure}

\subsubsection{$M_{\rm BH} - \sigma_{*}$}
\label{sec:M_BH_sigma}
Strong emission lines of AGNs may influence the accurate measurements of stellar velocity dispersion from absorption lines. \citet{oh11} tried to measure the velocity dispersion by masking the spectral regions potentially affected by nebular emissions, thus providing reliable velocity dispersion for emission-line galaxies in the OSSY catalog. Fig.~\ref{fig:velocity_disp} presents the relation between velocity dispersion and the magnitudes of galaxies. The format is the same as that in Fig.~\ref{fig:eff_radius}. We use the velocity dispersion given by the OSSY catalog and perform an aperture correction using the formula in \citet{cap06} to obtain the velocity dispersion at the effective radius $\sigma_{\rm e}$,
\begin{align}
\bigg(\frac{\sigma_{\rm aper}}{\sigma_{\rm e}}\bigg) = \bigg(\frac{r_{\rm aper}}{r_{\rm e}}\bigg)^{-0.066 \pm 0.035},
\label{eqn:aperture}
\end{align}
where $r_{\rm aper}$ is the aperture size of the SDSS fiber (1.5$\arcsec$ in radius) and $\sigma_{\rm aper}$ is the velocity dispersion observed through the fiber. Because of the inaccurate effective radius of the bulge for type 1 AGNs shown in Fig.~\ref{fig:eff_radius}, we instead use the galaxy effective radius. The correction factor to the velocity dispersion is small, and hence, the approximation has only a minor effect on our result.

Unlike the effective radius, the median velocity dispersion for type 1 (blue) and type 2 AGNs (red) is consistent especially when $M_{\rm r} \geq -22$. For $M_{\rm r} < -22$, the difference between the two types increases, as is clearly seen in the contours. The difference may be because broad components in the Balmer emission lines make it difficult to measure the velocity dispersion correctly.

\begin{figure}
\centerline{\includegraphics[width=8.5cm, angle=270]{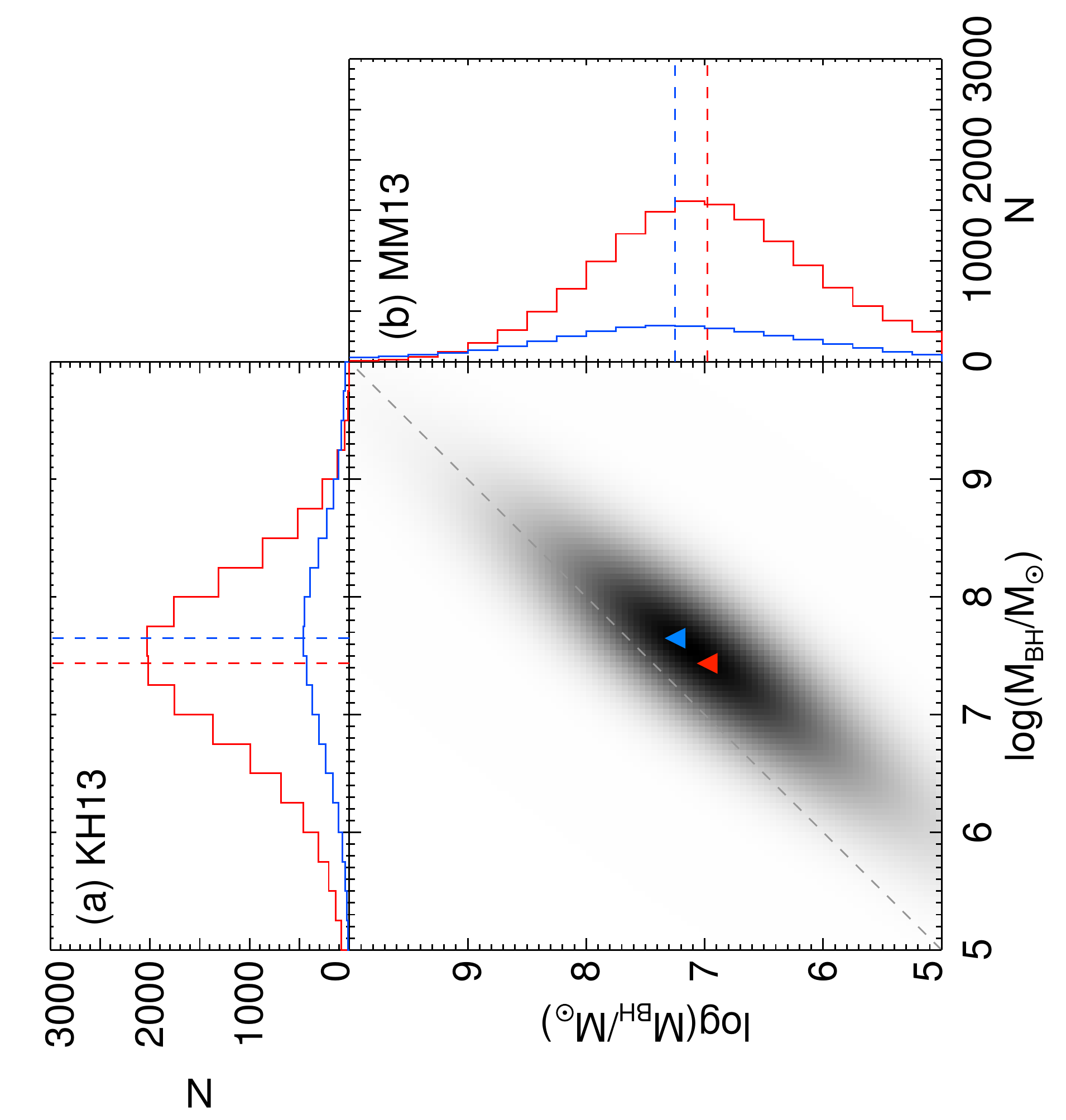}}
\caption[Distribution of the black hole mass derived from different $M_{\rm BH}-\sigma_*$ relations]{Distributions of black hole mass derived from different $M_{\rm BH}-\sigma_*$ relations and their comparison in the graycolor diagram. Each histogram shows the distribution of $M_{\rm BH}$ estimated from the relations of \citet{kor13} (panel (a)) and \citet{mcc13} (panel (b)). Type 1 and type 2 AGNs are represented in blue and red, respectively. The mean black hole mass for each type is indicated by dashed lines for histogram diagrams and by triangles for the gray-color plot. The gray dashed line indicate the one-to-one relation.}
\label{fig:M_BH_sigma}
\end{figure}

We derive $\MBH$ using the tight correlation between the supermassive black hole mass and the velocity dispersion given as
\begin{align}
\log \bigg(\frac{M_{\rm BH}}{M_\odot}\bigg) &= \alpha+\beta \log \bigg(\frac{\sigma_e}{200 \kms}\bigg).
\label{eqn:M_BH_sigma}
\end{align}
We use two recent $M_{\rm BH}-\sigma_*$ relations: $\alpha=8.49\pm0.05$ and $\beta=4.38\pm0.29$ with an intrinsic scatter of $\epsilon=0.29$ from \citet{kor13} and $\alpha=8.32\pm0.05$ and $\beta=5.64\pm0.29$ with $\epsilon=0.38$ from \citet{mcc13}. The steeper slope and the slightly large intrinsic scatter of \citet{mcc13} may be because they consider both classical and pseudo-bulges in deriving the $M_{\rm BH}-\sigma_*$ relation. Meanwhile, \citet{kor13} excluded pseudo-bulges, which generally have small $\MBH$ values and are scattered in the $M_{\rm BH}-\sigma_*$ plane.

To determine whether the relation of \citet{kor13} is applicable to our analysis, we estimate the number of pseudo-bulges in our sample. The S\'{e}rsic index is often used to discriminate pseudo-bulges from the classical bulges \citep{kor04,fis08}. The method, however, is not suitable for our sample, since the quality of the SDSS images for objects at $z>0.1$ is insufficient to obtain reliable S\'{e}rsic indices from decompositions. Instead, we use the bulge-to-total ratio in the $r$-band $(B/T)_{\rm r}\lesssim 0.2$ \citep{fis08,gad09} to select pseudo-bulges. For both types, only about 10\% of AGNs have $(B/T)_{\rm r}$ values smaller than $0.2$. Hence, the proportion of pseudo-bulges in our sample may not be large enough to affect our analysis when we utilize the relation of \citet{kor13}.
  
The distributions of $\MBH$ for the two methods are shown in Fig.~\ref{fig:M_BH_sigma}. Histograms indicate the distribution of the $\MBH$ estimated using the relation in \citet{kor13} (panel (a)) and that in \citet{mcc13} (panel (b)). Type 1 and type 2 AGNs are presented in blue and red, respectively. We take into account the intrinsic scatter of the $M_{\rm BH}-\sigma_*$ relations in calculating the mass distributions assuming a log-normal distribution of $\MBH$ at a given observed velocity dispersion \citep[e.g.,][]{mar04}, 
\begin{align}
&P(\log M_{\rm BH} | \log \sigma_*) \notag \\
&=  \frac{1}{\sqrt{2\pi}\epsilon}\exp\bigg[{-\frac{1}{2}\bigg(\frac{\log M_{\rm BH}-\alpha-\beta\log\sigma_*}{\epsilon}\bigg)^2}\bigg].
\label{eqn:intrinsic_scatter_consider}
\end{align}

\begin{figure}
\centerline{\includegraphics[width=8.2cm, angle=270, trim=0 0.5cm 0 0, clip]{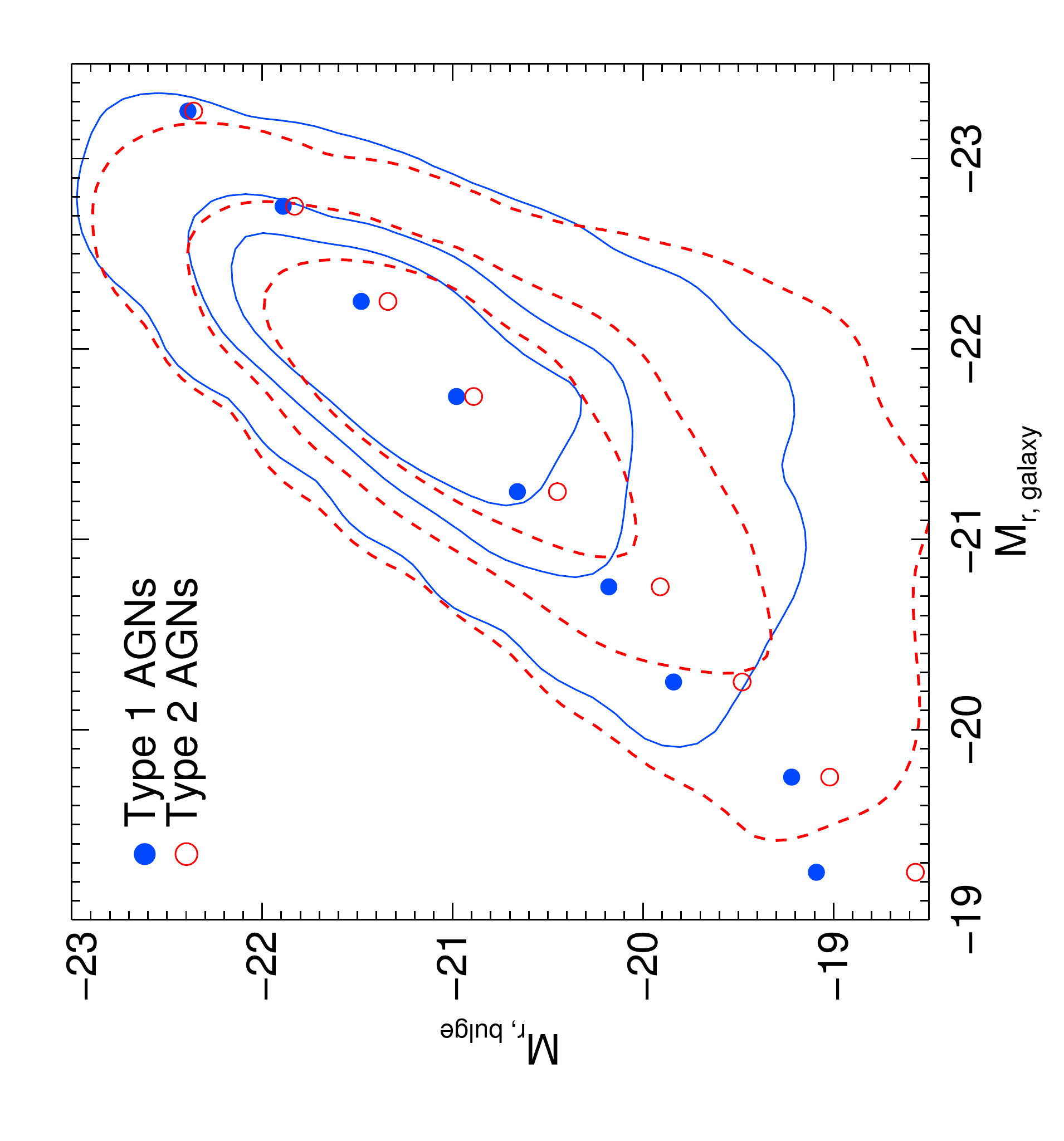}}
\caption[The bulge magnitude against the galaxy magnitude for type 1 and type 2 AGNs]{Bulge magnitude vs. galaxy magnitude for type 1 and type 2 AGNs. The format is the same as that in Fig~\ref{fig:eff_radius}.}
\label{fig:bulge_mag}
\end{figure}

Mean black hole masses (dashed lines) for type 1 AGNs are greater than those for type 2 AGNs in both panels. In addition, the representative values of type 1 and type 2 AGNs in panel (a) are greater than those in panel (b) by $\sim$$0.5$ dex. This is more clearly shown by the grayscale diagram in Fig.~\ref{fig:M_BH_sigma}. For most AGNs, $\MBH$ calculated based on the result of \citet{kor13} is larger than that based on \citet{mcc13}. In particular, 17\% of galaxies in panel (b) have $\MBH$ values smaller than $10^{6} M_{\odot}$, while only about 6\% of galaxies do so in panel (a). Because the difference is large enough to influence the trend of type 1 AGN fractions, we further compare the result of the $M_{\rm BH}-\sigma_*$ relation with $M_{\rm BH}$ from other methods in section~\ref{sec:M_BH_Halpha}.


\subsubsection{$M_{\rm BH}-L_{\rm bulge}$}
\label{sec:M_BH_luminosity}

\begin{figure}
\centerline{\includegraphics[width=8.5cm, angle=270]{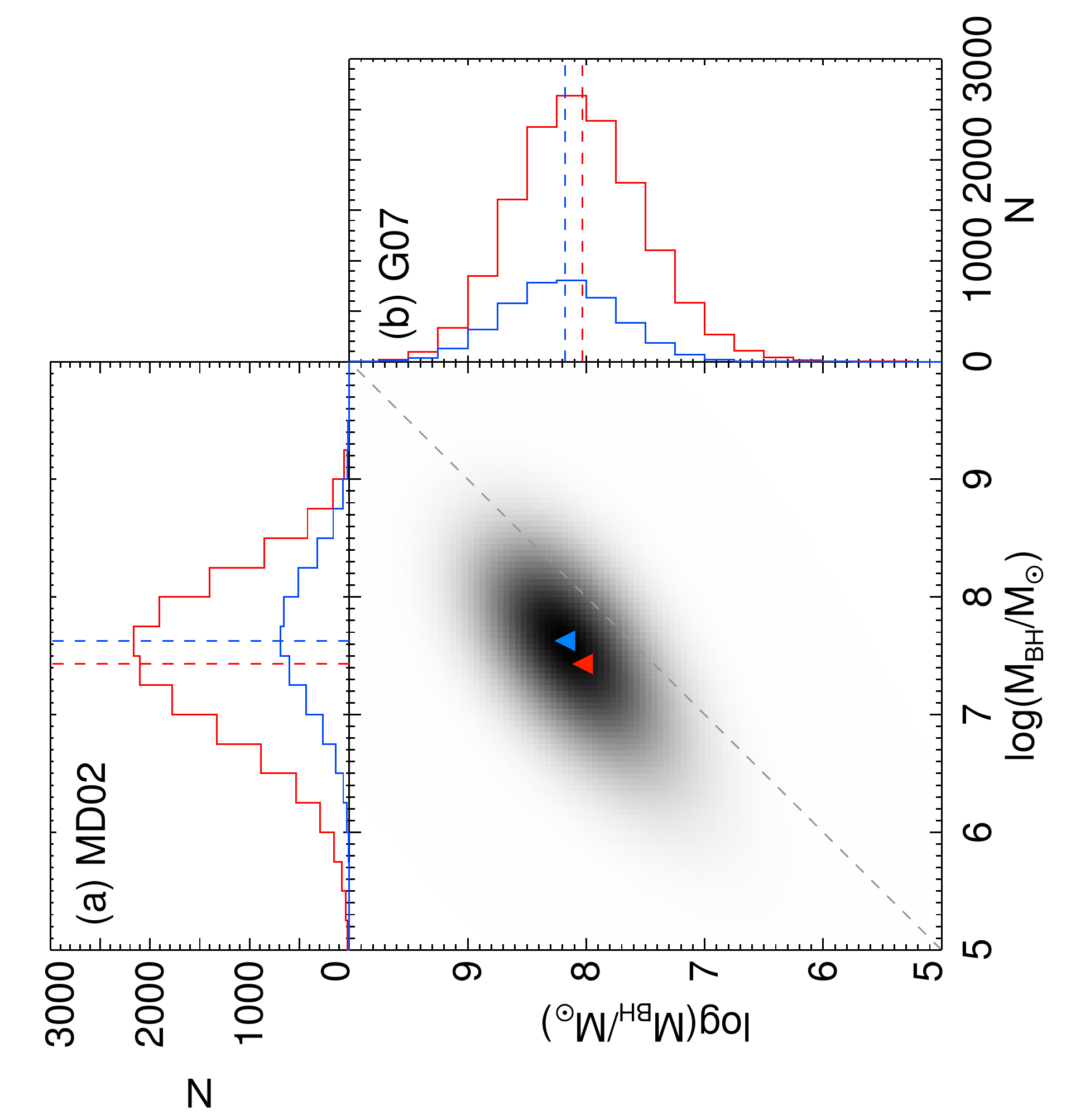}}
\caption[Distribution of the black hole mass derived from different $M_{\rm BH}-L_{\rm bulge}$ relations]{Distributions of black hole mass derived from different $M_{\rm BH}-L_{\rm bulge}$ relations and their comparison in a grayscale diagram. Each histogram shows the distribution of $M_{\rm BH}$ estimated from the relations of \citet{mcl02} (panel (a)) and \citet{gra07} (panel (b)). The format is the same as that in Fig.~\ref{fig:M_BH_sigma}.}
\label{fig:M_BH_luminosity}
\end{figure}

Bulge luminosity is regarded as a more robust measurement than other structural parameters of the bulge such as effective radius and S\'{e}rsic index \citep[e.g.,][]{ho14}. However, as in the case of the effective radius, bulge luminosity can also be affected by AGN contributions particularly for type 1 AGNs. Thus we investigate whether there is a difference between the bulge luminosities for type 1 and type 2 AGNs. Fig.~\ref{fig:bulge_mag} shows the bulge magnitudes in the $r$-band for type 1 and type 2 AGNs as functions of the magnitudes of the galaxies. The bulge magnitude is defined by \citet{sim11} and the galaxy magnitude is from the SDSS. Both magnitudes are extinction- and $k$-corrected. The format is the same as that in Fig.~\ref{fig:eff_radius}. Though the contours of type 2 AGNs are further extended to the lower luminosity than those of type 1 AGNs, the contour trends of the two types are almost same. Systematic offsets are observed in the median bulge magnitudes of type 1 and type 2 AGNs, but the difference between the two types does not prominently affect the derivation of the black hole mass through $M_{\rm BH}-L_{\rm bulge}$ relations.

\begin{figure*}[t]
\centerline{\includegraphics[width=8cm, angle=270, trim=0 0.5cm 0 0, clip]{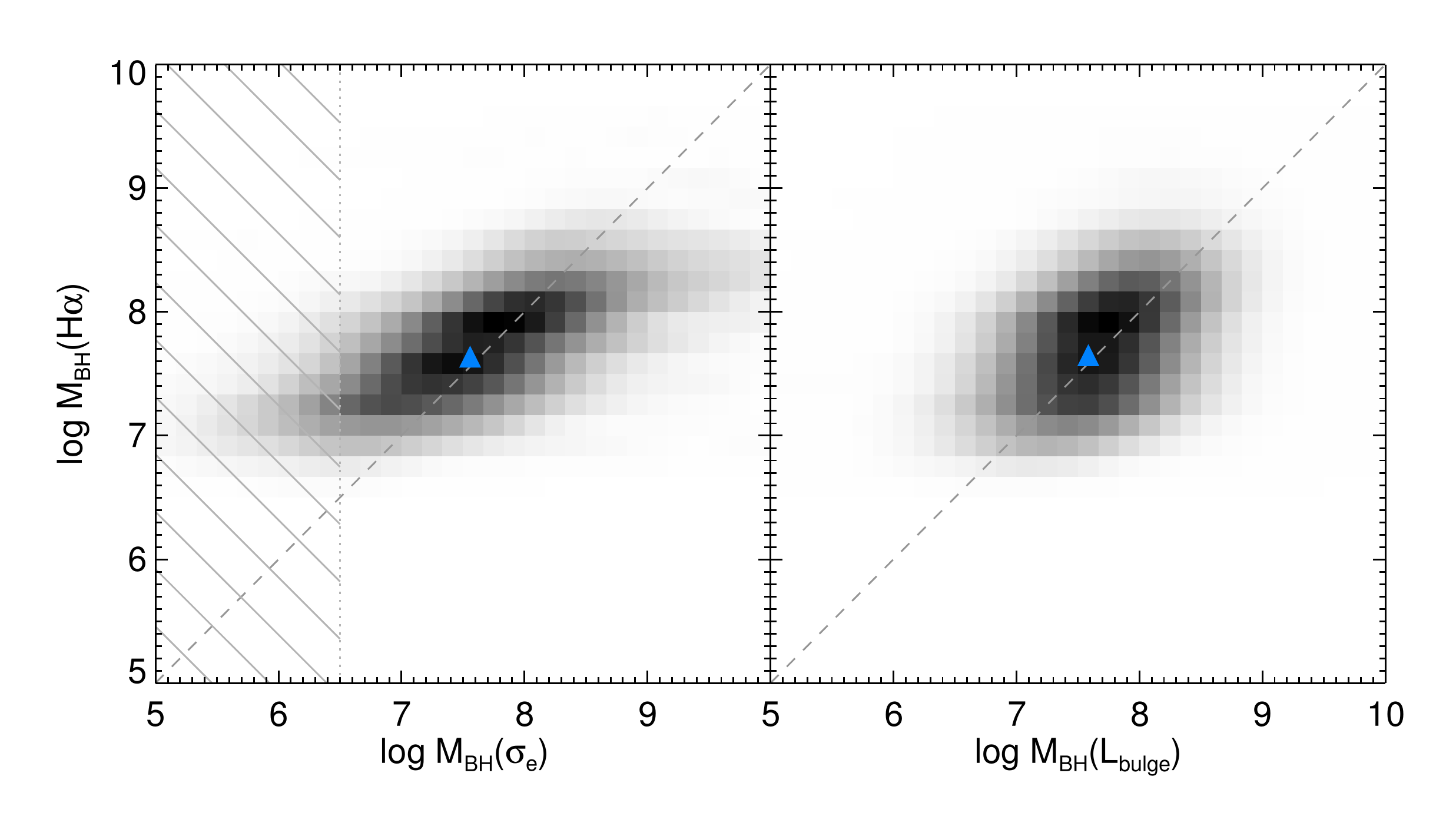}}
\caption[Reverberation Mapping]{Comparisons of the black hole masses derived from various methods. The $M_{\rm BH}$ estimated using equation~(\ref{eqn:M_BH_Halpha}) is presented on the y-axis and those from the $M_{\rm BH}-\sigma_*$ relation (left panel) and the $M_{\rm BH}-L_{\rm bulge}$ relation (right panel) are given on the x-axis. Both axes have units of solar masses. The blue triangle in each panel indicates the mean $\MBH$.}
\label{fig:M_BH_Halpha}
\end{figure*}

Assuming that $r-R\approx0.24$ \citep{fuk95} for galaxies, we calculate the black hole mass using two $M_{\rm BH}-L_{\rm bulge}$ relations,
\begin{equation}
\log \bigg(\frac{M_{\rm BH}}{M_\odot}\bigg) = -2.96\pm0.48-(0.50\pm0.02) M_{\rm R, bulge} ,
\label{eqn:MD02}
\end{equation}
\begin{equation}
\log \bigg(\frac{M_{\rm BH}}{M_\odot}\bigg) = 8.12\pm0.08-(0.38\pm0.04) (M_{\rm R, bulge}+21),
\label{eqn:G07}
\end{equation}
where equation~(\ref{eqn:MD02}) with an intrinsic scatter of $\epsilon=0.39$ is taken from \citet{mcl02} and equation~(\ref{eqn:G07}) with $\epsilon=0.30$ is from \citet{gra07}. \citet{gra07} derives equation~(\ref{eqn:G07}) based on the sample of \citet{mar03} but also obtains a similar result based on the sample of \citet{mcl02}. According to this study, different relations are obtained using the methods of \citet{mcl02} and \citet{gra07} for the same sample because the former uses $H_0=73 \rm \, km \, s^{-1} \, Mpc^{-1}$  while the latter uses $H_0=50 \rm \, km \, s^{-1} \, Mpc^{-1}$ to calculate absolute magnitudes.

The distributions of $M_{\rm BH}$ for the two methods are shown in Fig~\ref{fig:M_BH_luminosity}. Panels (a) and (b) show the $M_{\rm BH}$ values from equations~(\ref{eqn:MD02}) and (\ref{eqn:G07}), respectively. The format is the same as that in Fig.~\ref{fig:M_BH_sigma}. We consider the intrinsic scatter following equation~(\ref{eqn:intrinsic_scatter_consider}) but at the given bulge luminosity instead of the velocity dispersion. The mean $M_{\rm BH}$ of type 1 AGNs is slightly greater than that of type 2 AGNs for both panels as with the results of $M_{\rm BH}-\sigma_*$ relations (Fig.~\ref{fig:M_BH_sigma}). One remarkable point is that the overall distribution of panel~(b) is weighted toward massive $M_{\rm BH}$ by $\sim$$0.5$ dex compared to panel~(a). The grayscale density plot also illustrates that the $M_{\rm BH}$ based on the relation of \citet{gra07} is greater than that based on \citet{mcl02} on average. Although \citet{gra07} uses the updated data to derive the scaling relation, it appears that there are too many massive black holes with $M_{\rm BH} >10^9 M_\odot$ for our sample compared to the local universe \citep[e.g.,][]{kor13}. In the following section, we further investigate which $M_{\rm BH}$ derivation methods are appropriate for our study with the single-epoch H$\alpha$-based $\MBH$ estimate for type 1 AGNs.


\subsubsection{Single-Epoch H$\alpha$-based black hole mass estimates for type 1 AGNs}
\label{sec:M_BH_Halpha}

For type 1 AGNs, the black hole mass can be estimated by assuming that the dynamics of gas clouds in the BLR are dominated by the gravity of the central supermassive black hole,
\begin{align}
M_{\rm BH} = f\frac{r_{\rm BLR}V_{\rm BLR}^2}{G},
\label{eqn:RM_BH_mass}
\end{align}
where $G$ is the gravitational constant, $r_{\rm BLR}$ is the size of the BLR, $V_{\rm BLR}$ is the velocity of BLR clouds measured from the broad-line dispersion and $f$ is a dimensionless scaling factor of order unity. The factor $f$ which is determined by the combined effects of the geometry, kinematics and inclination of the BLR has not yet been confirmed and differs from study to study \citep[e.g.,][]{onk04,woo10,gra11,par12,gri13,ho14}. The BLR size $r_{\rm BLR}$ is estimated through the reverberation-mapping (RM) technique which measures the time lag between the variation of continuum of the compact accretion disk and that of emission lines of the BLR observed in multi-epoch spectra of AGNs \citep[e.g.][]{bla82,gas86,pet93}.

The black hole mass derivation method is further developed using the relation between the size of the BLR and the luminosity of AGN, which makes it possible to derive $\MBH$ by means of a single-epoch spectrum \citep[e.g.,][]{kas00,ben13}. The size-luminosity relation of AGNs is often constructed based on the continuum luminosity at $5100$\AA. \citet{gre05} found a tight relation between $L_{5100}$ and the luminosity of the H$\alpha$ emission line, allowing us to compute the BLR size with $L_{\rm H\alpha}$. Following the method used in \citet{gre05} and the size-luminosity relation of \citet{ben13}, \citet{rei13} present a black hole mass estimator as follows,
\begin{align}
\log \bigg(\frac{M_{\rm BH}}{M_\odot}\bigg) &= \log f + 6.57 + 0.47\log\bigg(\frac{L_{\rm H\alpha}}{10^{42} \ergs}\bigg) \notag \\ 
& \quad + 2.06\log\bigg(\frac{FWHM_{\rm H\alpha}}{10^3 \kms}\bigg),
\label{eqn:M_BH_Halpha}
\end{align}
where $L_{\rm H\alpha}$ is the luminosity of combined broad and narrow components of H$\alpha$ and $FWHM_{\rm H\alpha}$ is the FWHM of the broad H$\alpha$ emission. Here, we adopt the factor $f$ of $1.075$ \citep{gri13} as described in \citet{rei15}.

Fig.~\ref{fig:M_BH_Halpha} compares the $\MBH$ values estimated from different methods for type 1 AGNs. The x-axis indicates the $\MBH$ derived using the $M_{\rm BH}-\sigma_*$ relation of \citet{kor13} (left) and the $M_{\rm BH}-L_{\rm bulge}$ relation of \citet{mcl02} (right) and the y-axis represents the $M_{\rm BH}$ derived from equation~(\ref{eqn:M_BH_Halpha}). We consider the intrinsic scatter for $\MBH$ on the x-axis following equation~(\ref{eqn:intrinsic_scatter_consider}) as described in sections~\ref{sec:M_BH_sigma} and~\ref{sec:M_BH_luminosity}. In the right panel, the $\MBH$ values from the two methods are consistent with scatter, and the mean $\MBH$ (blue triangle) lies on the one-to-one line (dashed line). In case of the $M_{\rm BH}-\sigma_*$ relation (left panel), the $\MBH$ on the x-axis is more widely distributed than that on the y-axis, displaying discrepancies in low and high-mass black holes. However, high-density regions and the mean $\MBH$ still lie on the one-to-one line, indicating that the $\MBH$ values for most of type 1 AGNs are in reasonable agreement for different derivation methods.

\begin{figure*}
\centerline{\includegraphics[width=10cm, angle=270, trim=0 0.0cm 0 0, clip]{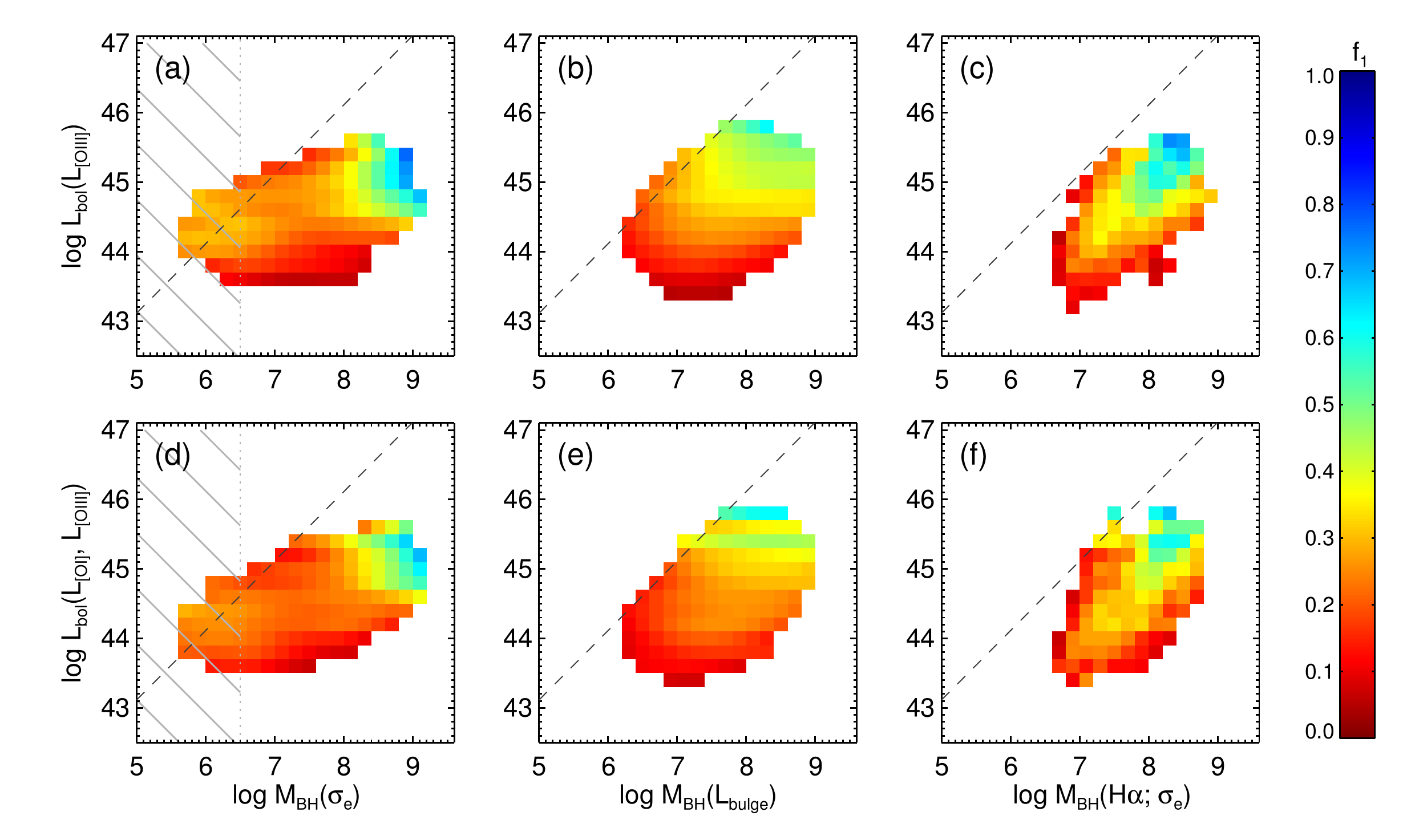}}
\caption[Type 1 AGN fractions on the $M_{\rm BH}-L_{\rm bol}$ planes]{Type 1 AGN fractions on the $M_{\rm BH}-L_{\rm bol}$ planes derived from observational data. The x-axis shows the $\MBH$ obtained from the $M_{\rm BH}-\sigma_*$ relation (left column), the $M_{\rm BH}-L_{\rm bulge}$ relation (middle column), and equation~(\ref{eqn:M_BH_Halpha}) for type 1 AGNs and the $M_{\rm BH}-\sigma_*$ relation for type 2 AGNs (right column). The y-axis shows $\Lbol$ estimated using $\OIII$ luminosity without reddening correction (top row) and the $\OI$ and $\OIII$ luminosities with reddening correction (bottom row). The x-axis is measured in solar masses and the y-axis is measured in $\ergs$. Each panel represents a combination of different derivation methods for $\MBH$ and $\Lbol$. Colors indicate the type 1 AGN fractions, with high fractions appearing in blue. The dashed line in each panel is the Eddington limit. The hatched region in the left column is where $\MBH$ is smaller than the resolution limit of the observation.}
\label{fig:frac_M_BH_L_bol}
\end{figure*}

The discrepancies at the high- and low-mass ends of the left panel may be caused by difficulty with measuring the velocity dispersion accurately. The spectroscopic resolution limit of the SDSS observation is $\sim$$70\kms$ for typical signal-to-noise ratios of the spectra \citep{ber03}, and thus the measurements below this limit are less reliable. The hatched region in Fig.~\ref{fig:M_BH_Halpha} indicates where $\MBH$ is smaller than the $\MBH$ corresponding to the limit ($\sim$$10^{6.5} M_\odot$). Assuming that the $\MBH$ given by equation~(\ref{eqn:M_BH_Halpha}) is more robust, $\MBH$ on the x-axis in the hatched region may be underestimated. Moreover, the velocity dispersion of type 1 AGNs tends to be overestimated for massive galaxies and black holes, as shown in Fig.~\ref{fig:velocity_disp}. As discussed in section~\ref{sec:M_BH_sigma}, this may be because the broad emissions of type 1 AGNs are much more prominent than the absorption lines particularly in the case of massive black holes, making it difficult to measure the stellar velocity dispersion accurately. Therefore, the differences of $\MBH$ at the low- and high-mass ends are likely to be partially attributable to unreliable velocity dispersion.

As discussed in sections~\ref{sec:M_BH_sigma} and~\ref{sec:M_BH_luminosity}, the $\MBH$ computed with the relation in \citet{mcc13} is smaller than that computed with the relation in \citet{kor13}, and the $\MBH$ from the relation of \citet{gra07} is larger than that from \citet{mcl02}. This means that based on the comparison with the single-epoch H$\alpha$-based $\MBH$ estimates, the relation of \citet{mcc13} underestimates $\MBH$ for our sampe and the relation of \citet{gra07} overestimates it. Therefore, in our analysis of the type 1 AGN fraction, we use the relations of \citet{kor13} and \citet{mcl02}, which provide more consistent $M_{\rm BH}$ estimates, and additionally consider equation~(\ref{eqn:M_BH_Halpha}) for type 1 AGNs. It is beyond the scope of our study to judge the accuracy of each method, and thus we first treat the three methods with equal weights.


\section{Fraction of Type 1 AGNs as a Function of $M_{\rm BH}$ and $L_{\rm bol}$}
\label{sec:fraction_2D}
In this section, we examine the behavior of the type 1 AGN fraction on the two-dimensional space defined by $\MBH$ and $\Lbol$. Since we consider two derivation methods for $\Lbol$ and three derivation methods for $\MBH$, there are six combined cases. Fig.~\ref{fig:frac_M_BH_L_bol} shows the type 1 AGN fractions for various $\MBH$ and $\Lbol$. Each panel corresponds to a different combination of the methods. The x-axis indicates the $\MBH$ estimated with the $M_{\rm BH}-\sigma_*$ relation of \citet{kor13} (left column; panels (a) and (d)) and with the $M_{\rm BH}-L_{\rm bulge}$ relation of \citet{mcl02} (middle column; panels (b) and (e)) with consideration of the intrinsic scatter (equation~(\ref{eqn:intrinsic_scatter_consider})). In the right column (panels (c) and (f)), we calculate the black hole mass by applying equation~(\ref{eqn:M_BH_Halpha}) for type 1 AGNs and the $M_{\rm BH}-\sigma_*$ relation of \citet{kor13} for type 2 AGNs. The y-axis shows $\Lbol$ computed with the relation of \citet{hec04} (top row; panels (a) and (b)) and with the relation of \citet{net09} (bottom row; panels (c) and (d)). Only grids with type 1 and type 2 AGNs greater than 5 are filled. Bluer colors indicate higher type 1 AGN fractions. Dashed lines indicate the Eddington limit.

In each column, the trends of the type 1 AGN fraction in the top and bottom panels seem similar because $\Lbol$ is not severely affected by the choice of derivation methods, as discussed in section~\ref{sec:L_bol}. However, $\MBH$ varies depending on what types of derivation methods are applied (section~\ref{sec:M_BH}), presenting different trends of the type 1 AGN fraction for different methods.

In the first column (panels (a) and (c)), the galaxies are distributed over wide ranges of $\MBH$ as shown in Fig.~\ref{fig:M_BH_sigma}, leading to some galaxies being above the Eddington limit. Most of those galaxies, however, are included in the hatched region where $\MBH$ is smaller than the $\MBH$ value corresponding to the spectroscopic limit of the SDSS (section~\ref{sec:M_BH_Halpha}), and the $\MBH$ in this region may be underestimated. Furthermore, the extent of the spread of $\MBH$ is different for type 1 and type 2 AGNs. The dispersion of the $\MBH$ distribution of type 1 AGNs is larger than that of type 2 AGNs, resulting in high type 1 AGN fractions at both small and large $\MBH$. Again, however, the fractions at $M_{\rm BH} < 10^{6.5} M_\odot$ may not be reliable since $\MBH$ below the limit is uncertain. The different dispersions, as well as different mean $\MBH$ values, between the two types also cause the variation in the type 1 AGN fraction along the x-axis to be more prominent than that along the y-axis.

In the second column of Fig.~\ref{fig:frac_M_BH_L_bol} (panels (b) and (d)), the galaxies cover narrower ranges of $M_{\rm BH}$ than those in the first column. As a result, there are only a small number of galaxies above the Eddington limit. Moreover, the fraction of type 1 AGNs at $M_{\rm BH} > 10^8 M_\odot$ is lower than that shown in the first column, whereas the fraction at the intermediate region ($10^7 < M_{\rm BH} < 10^8 M_\odot$) is higher than that in the first panel. This may be because, unlike the $M_{\rm BH}$ distribution of the $M_{\rm BH}-\sigma_*$ relation mentioned above, the distributions derived from the $M_{\rm BH}-L_{\rm bulge}$ relation for type 1 and type 2 AGNs have similar dispersions with a small offset. Consequently, the fraction of type 1 AGNs changes mildly along $\MBH$ and seems more affected by $\Lbol$.

When deriving $\MBH$, \citet{oh15} chose the single-epoch $\MBH$ estimate devised by \citet{gre05} for type 1 AGNs and the $M_{\rm BH}-\sigma_*$ relation of \citet{gul09} for type 2 AGNs. Based on the choice, \citet{oh15} found a ridge-shaped distribution of the type 1 AGN fraction on the $\MBH-\Lbol$ plane, which shows an increasing and then decreasing trend of the fraction along $\MBH$ or $\Lbol$. This ridge-shaped feature is not clearly observed in the first and second columns. In the third column (panels (c) and (f)), however, the ridge-shaped distribution is reproduced when we follow the approach of \citet{oh15}, but with updated relations. The fraction changes similarly along both the x- and y-axes, and seems equally affected by $\MBH$ and $\Lbol$.

Despite differences in detail, the general trends of the fraction are similar for all panels, showing the higher type 1 AGN fractions observed at large $\MBH$ and high $\Lbol$. The changes in the fraction along $\MBH$ at a fixed $\Lbol$ and along $\Lbol$ at a fixed $\MBH$ indicate that $\MBH$ and $\Lbol$ play separate roles in forming the structure of the torus. In order to investigate the dynamic system of the AGN torus, several studies have used hydrodynamic simulations \citep[e.g.,][]{dor11,hop12,wad12}. For example, using multiscale simulations, \citet{hop12} found that gravitational instabilities can generate thick and torus-sized disks and explain the decreasing trend of the type 2 AGN fraction (or the increasing trend of the type 1 AGN fraction) with $\MBH$. In addition, \citet{liu11} developed a torus model based on analytic calculations with anisotropic radiation from the accretion disk to explain the type 2 AGN fraction, but they did not consider the scheme of the clumpy torus model. In this study, we suggest an analytic torus model based on the clumpy-torus scheme to examine the trends of the fraction, and the structure of the AGN torus, as a function of $\MBH$ and $\Lbol$.


\begin{figure*}
\centerline{\includegraphics[width=13.5cm, trim=0 0.0cm 0 0, clip]{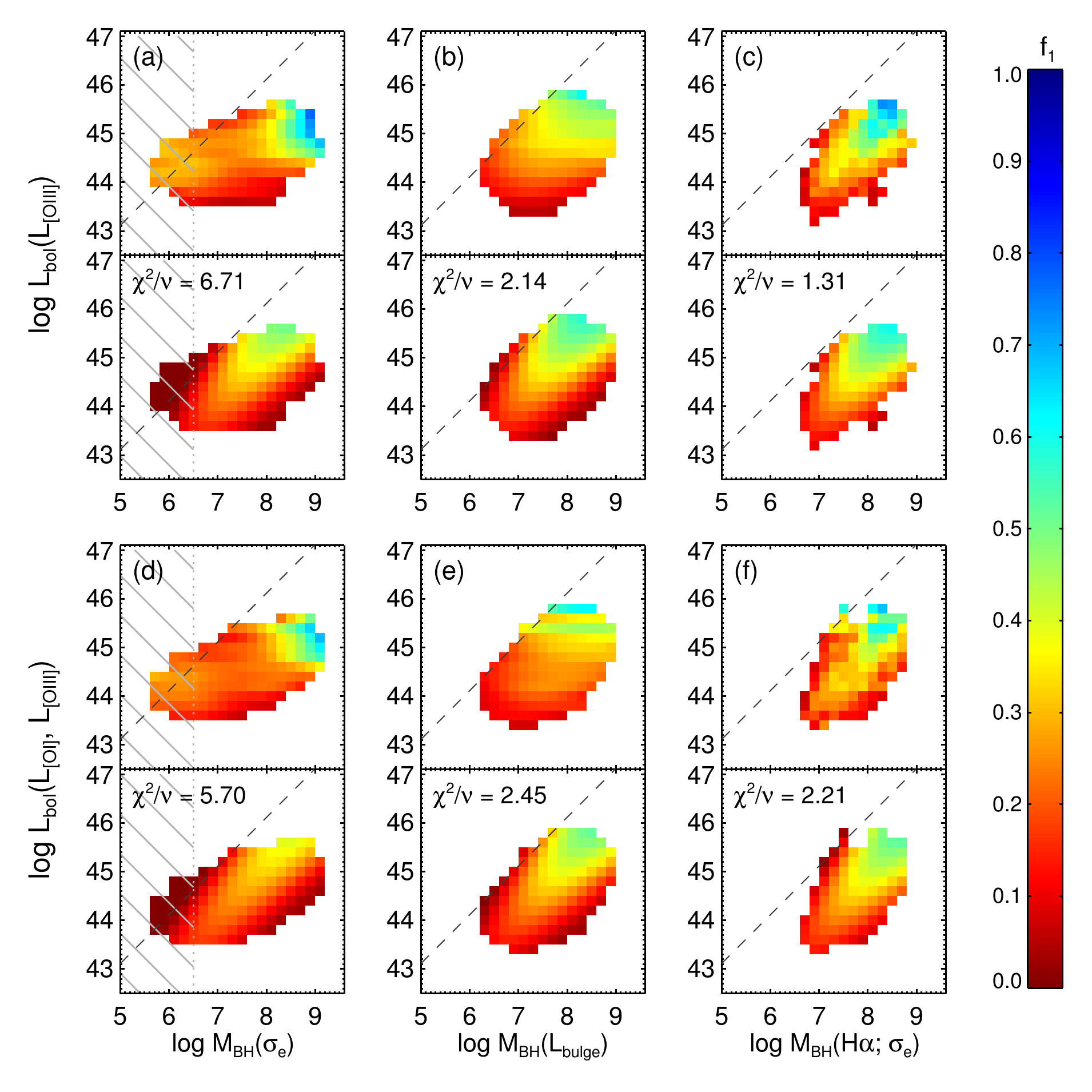}}
\caption[Results of the fit using the clumpy torus model]{Results of the fit using the analytic torus model. Each pair shows the observation (top) and the model fit (bottom). The dashed line in each panel is the Eddington limit. The notations from (a) to (f) are the same as those in Fig.~\ref{fig:frac_M_BH_L_bol}. $\chi^2_\nu$ is indicated at the top-left corner of each bottom panel.}
\label{fig:frac_clumpy}
\end{figure*}

\section{Discussion}
\label{sec:discussion}

The torus is thought to be composed of a number of clumpy clouds rather than a smooth and continuous dust distribution \citep{kro88,nen02,nen08a,nen08b}. In the clumpy cloud scheme, the photons emitted from the BLR can penetrate the torus through the gaps between the clouds. Therefore, unlike the smooth torus model, the fraction of type 1 AGNs is determined not only by the geometry of the torus but also by the number of clumpy clouds. In this section, we suggest a torus model based on the formalism of \citet{nen02,nen08a,nen08b} to explain the behavior of the type 1 AGN fraction shown in Fig.~\ref{fig:frac_M_BH_L_bol}.

In the clumpy cloud model with optically-thick clouds, the probability of seeing the center passing through the torus is determined by the number of clumpy clouds along the line-of-sight, $P_{\rm esc}(\beta) = \exp(-N(\beta))$ where $\beta$ is the angle between the line-of-sight and the equatorial plane. Assuming a Gaussian angular distribution, the number of clumpy clouds in the line-of-sight can be expressed as $N(\beta) = N_{\rm 0}\exp (-(\frac{\beta}{\sigma})^2)$ \citep{nen02}. Here, $\sigma$ is the angular width of the torus measured from the equatorial plane and $N_0$ is the number of clouds along a radial equatorial ray. By integrating the probability over the solid angle, we can obtain the type 1 AGN fraction,
\begin{align}
f_{\rm 1} &= \frac{1}{4\pi}\int P_{\rm esc}(\beta) \, d \Omega \notag \\
\, &= \int^{\pi/2}_{0} \exp(-N(\beta)) \cos(\beta) \, d \beta.
\label{eqn:frac_clumpy}
\end{align}
\noindent This fraction is determined by the number of clouds along a radial equatorial ray ($N_0$) and the angular width of the torus ($\sigma$).

To understand the structural parameters of the torus, we assume that its angular width is mainly determined by a tug-of-war between the radiation pressure and the gravitational force on the clumpy clouds. Here, we do not take into account a thermal-driven wind \citep[e.g.,][]{eve07}, a magnetohydrodynamic wind \citep[e.g.,][]{eve05} or turbulence, although they can influence the dynamics of the clouds. Instead, we focus on the radiation pressure for simplicity. An anisotropic radiation emitted from the accretion disk is strongest in the direction the normal to the accretion disk ($\beta=90^\circ$), becoming weaker as $\beta$ approaches $0^\circ$. This means that the clouds at the angle where radiative force is stronger than the gravitational force will be pushed away; otherwise, the clouds would remain, forming a torus. Thus, considering the radiative force coupled with the gravitational force on the clouds, we calculate the critical angle at which the two forces are balanced.

\begin{figure*}
\centerline{\includegraphics[width=5.2cm, angle=270, trim=0 0.0cm 0 0, clip]{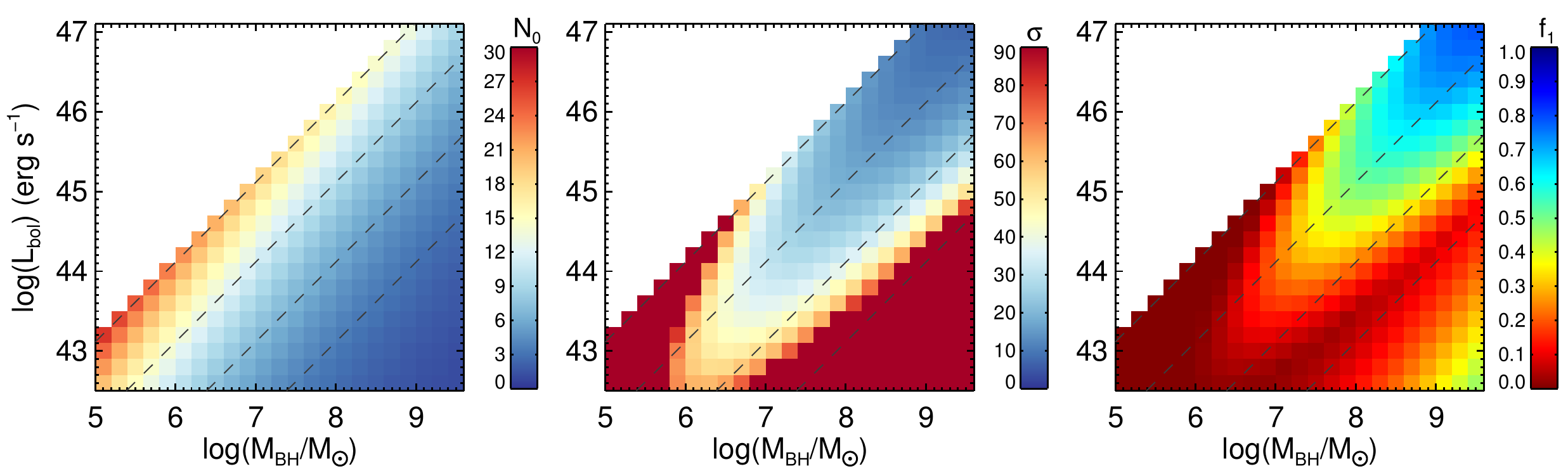}}
\caption[$N_0$ and $\sigma$ from the clumpy torus model]{Behaviors of $N_0$ (left), $\sigma$ (middle), and the type 1 AGN fraction from the model (right) with respect to $M_{\rm BH}$ and $L_{\rm bol}$. The levels of the quantities are indicated by the color bars. The dashed lines in each panel correspond to $L_{\rm bol} = L_{\rm Edd}$, $0.1L_{\rm Edd}$, $0.01L_{\rm Edd}$ and $0.001L_{\rm Edd}$ from top to bottom.}
\label{fig:model_N0_sigma_frac}
\end{figure*}

The acceleration of the radiation-pressure-driven motion is defined as
\begin{align}
a_{\rm rad} = \frac{\kappa F}{c},
\end{align}
where $\kappa$ is the opacity, $c$ is the speed of light and $F$ is the flux from the accretion disk. \citet{sun85} suggest the anisotropic radiation in the form of
\begin{align}
F(r,\beta)=\frac{3L}{14\pi r^2}\sin\beta\,(1+2\sin\beta),
\end{align}
where $L$ is the bolometric luminosity and $r$ is the distance from the center to a cloud. With the opacity $\kappa = \tau_{\rm V}/(m_{\rm p}N_{\rm H})$, where $\tau_{\rm V}$ is the optical depth at visual, $m_{\rm p}$ is the proton mass and $N_{\rm H}$ is the column density of the cloud, the acceleration is
\begin{align}
a_{\rm rad}(r,\beta) &= \frac{3L}{14\pi m_{\rm p}cr^2}\frac{\tau_{\rm V}}{N_{\rm H}}\sin\beta\,(1+2\sin\beta).
\end{align}
At the outer part of the torus where clouds can escape from their reservoir, the clouds may be shielded by other clouds in the same ray. Considering the shielding effect by multiplying a factor of $e^{-N(\beta)}$, we obtain the acceleration at the outer radius ($r_{\rm out}$) as follows:
\begin{align}
a_{\rm rad}(\beta) = 1.40\times 10^{-5} \frac{\sin\beta\,(1+2\sin\beta)\,L_{45}}{(r_{\rm out}/\rm pc)^2}\frac{\tau_{\rm V}}{N_{23}}\,e^{-N(\beta)},
\end{align}
where $L_{45}=L/10^{45} \ergs$ and $N_{23}=N_{\rm H}/10^{23} \,\rm cm^{-2}$. The acceleration due to gravity at the outer region is simply given by
\begin{align}
a_{\rm grav} = \frac{GM_{\rm BH}}{r_{\rm out}^2} = 1.39\times 10^{-5} \frac{M_8}{(r_{\rm out}/\rm pc)^2},
\end{align}
where $M_8 = M_{\rm BH}/10^8 M_\odot$. Assuming $\beta=\sigma$ in the mechanical equilibrium ($a_{\rm rad}=a_{\rm grav}$), the angular width of the torus can be described as
\begin{align}
\sin\sigma = -0.25+0.25\,\sqrt{1+8\,\frac{N_{23}}{\tau_{\rm V}}\frac{M_8}{L_{45}}\,e^{N_0/e}}.
\label{eqn:sigma}
\end{align}

There is no exact function of $N_0$ with respect to $\MBH$ and $\Lbol$, and either quantity may have positive or negative effects on $N_0$. For instance, high luminosities in AGNs can be triggered by large amounts of gas infall, which may supply gas and dust to the torus (a positive relation between $N_0$ and $\Lbol$), but at the same time, the strong radiation can remove the surrounding materials by evaporating or blowing them away (a negative relation between $N_0$ and $\Lbol$). Furthermore, it is not straightforward to predict the relation between $\MBH$ and $N_0$. There are studies indicating that stellar bars, as an internal mechanism \citep[e.g.,][]{oh12,gal15}, and galaxy-galaxy interactions, as an external mechanism \citep[e.g.,][]{kau04,pim13}, can play a role in driving gas inflows and triggering AGN activities by exerting gravitational torques. In those studies, massive galaxies hosting massive black holes, on average, are more likely to show signs of AGN activity than low-mass galaxies. However, the observed high frequency of AGNs at massive black holes does not mean that the amount of infalling gas (or $N_0$) is large for massive black holes and the effect of $\MBH$ on $N_0$ is not clear. In this study, we simply assume that $N_0$ is determined by $\MBH$ and $\Lbol$ as,
\begin{align}
N_{\rm 0} &= \tilde{N_{\rm 0}} \,(M_8)^a (L_{45})^b,
\label{eqn:N0}
\end{align}
where $\tilde{N_{\rm 0}}$, $a$ and $b$ are free parameters.

\begin{table}[t] {
\caption[Fitting parameters of the clumpy torus model]{Fitting parameters of the torus model for the six panels in Fig.~\ref{fig:frac_clumpy}.}
\label{tab:fitting}
\begin{center} {
\begin{tabular}{C{0.7cm}  | C{1.1cm} | C{1.1cm} C{1.1cm} C{1.1cm}}
\hline
\hline \\[-0.21cm]
	&	$\chi_\nu^2$	&	$\tilde{N_{\rm 0}}$	 &	$a$		&	$b$ \\[0.03cm]
\hline \\[-0.21cm]
(a)	&	6.71	&	9.6	&	-0.25 	&	0.20	\\[0.02cm]
(b)	&	2.14	&	9.4	&	-0.25 	&	0.19	\\[0.02cm]
(c)	&	1.31	&	9.0	&	-0.30 	&	0.22  	\\[0.02cm]		
(d)	&	5.70	&	10.1	&	-0.24 	&	0.21  	\\[0.02cm]
(e)	&	2.45	&	9.7	&	-0.25		&	0.20	\\[0.02cm]
(f)	&	2.21	&	9.5	&	-0.29 	&	0.23	\\
\hline
\end{tabular}}
\end{center}
}
\end{table}

\begin{figure*}
\centerline{\includegraphics[width=18cm, trim=0 1.3cm 0 0, clip]{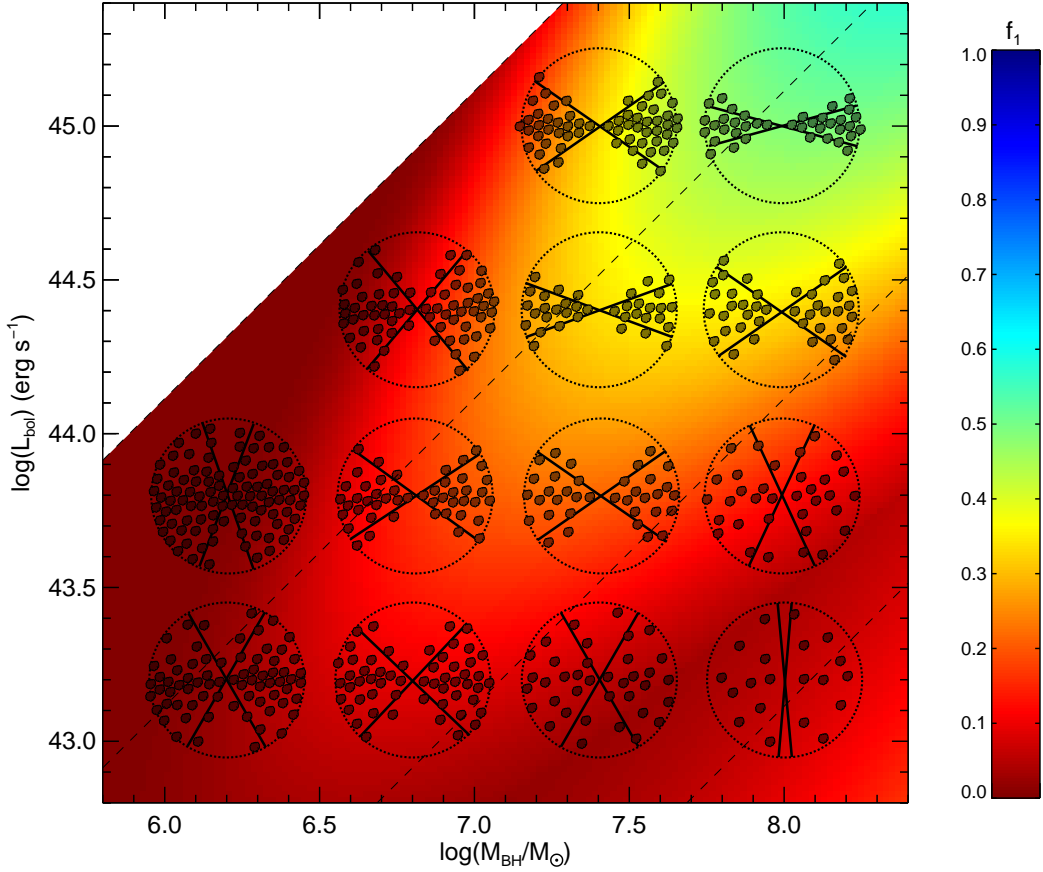}}
\caption[Schematic Diagram]{The schematic diagram showing the type 1 AGN fraction determined by $N_0$ and $\sigma$. Dotted circles indicate torus models with different $\MBH$ and $\Lbol$. Solid lines in each model mean the angular width of the torus and the symbols are clouds. The trend of the fraction is presented by the background color. Dashed lines correspond to $L_{\rm bol} = L_{\rm Edd}$, $0.1L_{\rm Edd}$, $0.01L_{\rm Edd}$ and $0.001L_{\rm Edd}$ from top to bottom. The number of clouds along a radial equatorial ray ($N_0$) gradually decreases with increasing $\MBH$ and increases with increasing $\Lbol$, resulting in large $N_0$ at $\EddRatio>0.1$. The angular width of the torus ($\sigma$) decreases and then increases again as $\Lbol$ or $\MBH$ increases at a fixed line. The large $\sigma$ at high luminosity or small black hole is caused by the shielding effect that makes radiative pressure difficult to blow away the clouds.}
\label{fig:schematic}
\end{figure*}

The results of fits using equation~(\ref{eqn:frac_clumpy}) with equations~(\ref{eqn:sigma}) and (\ref{eqn:N0}) are shown in Fig.~\ref{fig:frac_clumpy}. We take $N_{23}=1$ and $\tau_{\rm V}=50$ which are regarded as typical values \citep[e.g.,][]{nen08b,mor09,min14}. Each panel pair represents the observation (top) and the best model fit (bottom). The reduced $\chi^2$ appears at the corner of each bottom panel. In all panels, model fits present ridge-shaped features. The observed trends of the type 1 AGN fraction based on the $M_{\rm BH}-\sigma_*$ relation (panels (a) and (d)), which do not show the ridge-shaped distribution, are poorly fitted and cannot be explained by the model. The results of the $M_{\rm BH}-L_{\rm bulge}$ relation (panels (b) and (e)) are fitted better than those of the $M_{\rm BH}-\sigma_*$ relation but the fits still do not seem to reproduce the observations properly, as the ridge-shaped features shown in the model fits are not obvious in the observations. In the last column, however, the ridge-shaped distribution of the model is well matched with the observation. The model fit in the panel (c) gives $\chi^2_\nu=1.31$ and well reproduce the observed trend. Though the $\chi^2_\nu$ value in panel (f) is rather large, the general trend of the fit follows the observation.

The parameters of best fit for each panel are given in Table~\ref{tab:fitting}. Although $\chi^2_\nu$ is poor in some panels, the best-fit parameters are almost consistent, regardless of the method used to calculate $\MBH$ and $\Lbol$. The negative exponent $a$ means that $N_0$ and $\MBH$ are anti-correlated, suggesting that there is small $N_0$ for large $\MBH$. Meanwhile, $N_0$ and $\Lbol$ are positively correlated, and thus $N_0$ increases with $\Lbol$.

Based on the best-fit result for panel (c), we explore the distribution of $N_0$ and $\sigma$ in the $\MBH-\Lbol$ plane. Fig.~\ref{fig:model_N0_sigma_frac} demonstrates the behaviors of $N_0$ (left), $\sigma$ (middle), and the fraction of type 1 AGNs (right) with respect to $\MBH$ and $\Lbol$. The levels of the quantities are indicated by the color bars. The dashed lines correspond to $\Lbol = L_{\rm Edd}$, $0.1L_{\rm Edd}$, $0.01L_{\rm Edd}$ and $0.001L_{\rm Edd}$ from top to bottom. We only examine the region where $\Lbol < L_{\rm Edd}$.

In the left panel of Fig.~\ref{fig:model_N0_sigma_frac}, $N_0$ shows monotonic trends against $\MBH$ and $\Lbol$; it decreases for $\MBH$ and increases for $\Lbol$. Nonzero $a$ and $b$ are critical in reproducing the ridge-shaped distribution of the observation, as $N_0$ without $\MBH$- and $\Lbol$-dependency cannot generate the trend of $\sigma$ shown in the middle panel. Previous studies in the mid-IR have investigated $N_0$ by fitting the spectra with the clumpy torus model \citep[e.g.,][]{mor09,ram09,ram11,alo11,ich15}, and found $N_0$ ranging from 1 to 15 for their AGN samples. The results are consistent with our results which show $N_0$ spanning from $1$ to $18$ for the same coverage of $\MBH$ as that used in the previous studies (i.e., $10^7<\MBH<10^9 M_\odot$).

In the case of the angular width $\sigma$ (middle panel), the trend is not monotonic. The effect of $\Lbol$ on $\sigma$ differs for different ranges of the Eddington ratio ($\lambda_{\rm Edd}=L_{\rm bol}/L_{\rm Edd}$). When $\lambda_{\rm Edd}<0.1$, $\sigma$ decreases along $\Lbol$ at a given $\MBH$. This is consistent with the expectation of the receding model that the opening angle of the torus (or the inverse of the angular width of the torus) increases with $\Lbol$. When $\lambda_{\rm Edd}>0.1$, however, the angular width at a given $\MBH$ increases as $\Lbol$ increases. This is because, for the torus with $\EddRatio<0.1$, there are a small number of clumpy clouds as shown in the left panel, and thus, the shielding effect is small. In other words, for $\EddRatio<0.1$, the clouds are easily affected by the radiative force and the angular width becomes thin with increasing luminosity. When $\EddRatio>0.1$, however, there are a large number of clouds, and thus, the radiative force is weakened by the shielding effect, making it difficult to push the clouds away. Consequently, despite increasing luminosity, $\sigma$ becomes thick again. Similarly, as $\MBH$ increases at a given $\Lbol$, $\sigma$ decreases, because of the shielding effect, and then turns into an increasing trend at around $\EddRatio=0.1$.

The type 1 AGN fraction in the right panel of Fig.~\ref{fig:model_N0_sigma_frac} can be explained by the combined effect of $N_0$ and $\sigma$. When $\EddRatio > 10^{-3}$, the general trend of the type 1 AGN fraction follows the trend of $\sigma$, and hence the fraction is primarily determined by $\sigma$ and marginally affected by $N_0$. When $\EddRatio < 10^{-3}$, $\sigma$ is $90^\circ$ but $N_0$ approaches $0$. As a result, the escape probability of photons emitted from the center becomes high and the type 1 AGN fraction increases. According to \citet{nic00}, however, the BLR cannot exist for the Eddington ratio lower than $\sim$$10^{-3}$. This means that, if we consider the absence of the BLR, the type 1 AGN fraction in the region where $\lambda_{\rm Edd} \lesssim 10^{-3}$ would be close to $0$, even though we can see the center through the torus.

\begin{figure}
\centerline{\includegraphics[width=8.5cm, trim=0 0.2cm 0 0, clip]{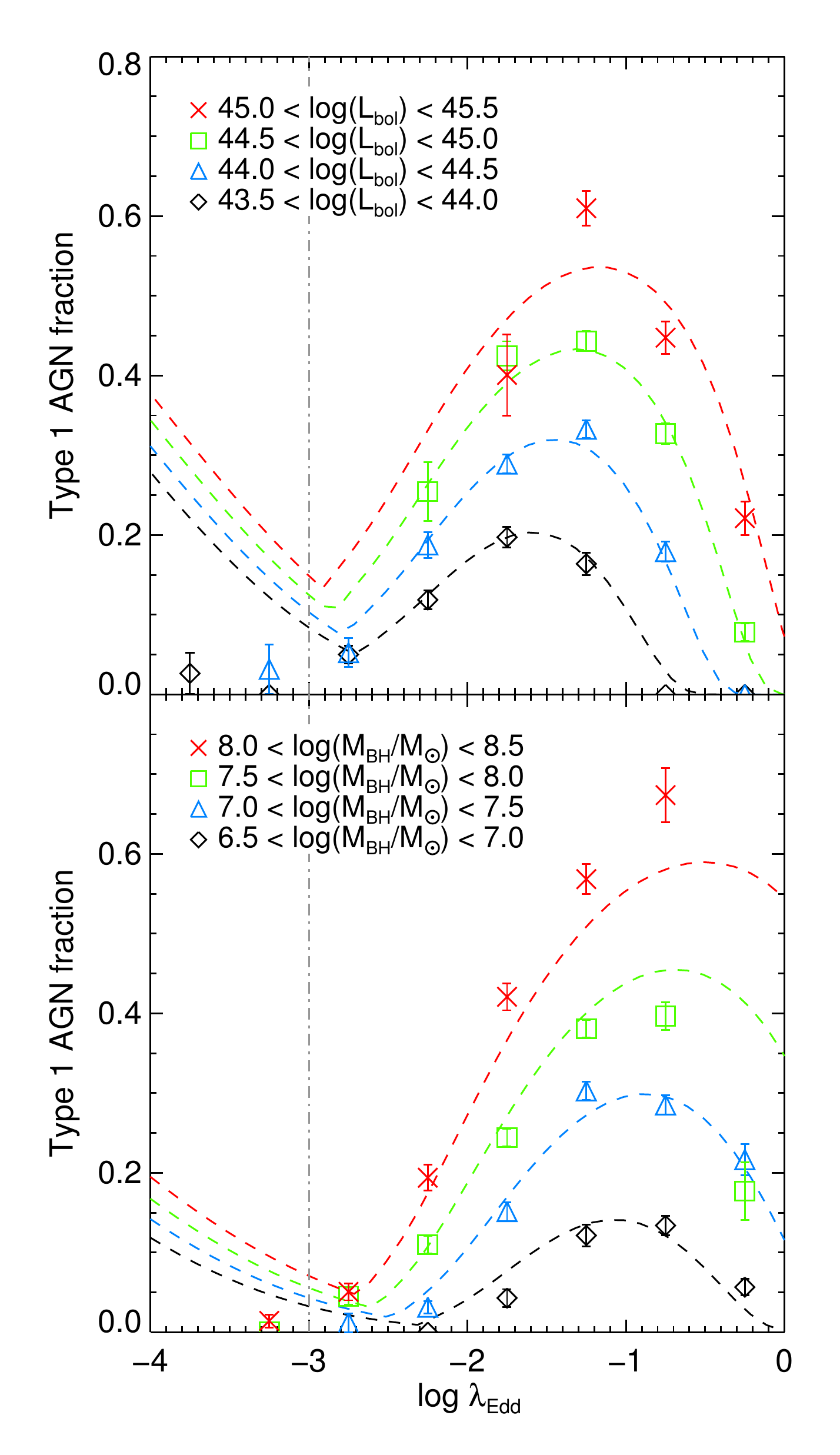}}
\caption[Type 1 AGN fraction along the Eddington ratio]{Top: the fraction of type 1 AGNs with respect to the Eddington ratio for different ranges of $L_{\rm bol}$. Bottom: the fraction of type 1 AGNs with respect to the Eddington ratio for different ranges of $M_{\rm BH}$. Different colors indicate different ranges. Symbols represent observations and dashed lines are from the model. Dot-dashed lines mean the boundary where the BLR cannot exist and thus the fraction would be close to 0.}
\label{fig:frac_Eddington_ratio}
\end{figure}

Fig.~\ref{fig:schematic} illustrates a schematic diagram that shows the effects of $N_0$ and $\sigma$ on the fraction of type 1 AGNs indicated by colors. Each dotted circle represents the torus model with different $\MBH$ and $\Lbol$ and black solid lines in each circle show the angular width of the torus. Dashed lines correspond to $L_{\rm bol} = L_{\rm Edd}$, $0.1L_{\rm Edd}$, $0.01L_{\rm Edd}$ and $0.001L_{\rm Edd}$ from top to bottom. The diagram presents that, as mentioned above, $N_0$ decreases with increasing $\MBH$ at a given $\Lbol$ and increases with increasing $\Lbol$ at a given $\MBH$, resulting in a large number of clouds at $\EddRatio>0.1$. As a result, at a given $\MBH$, the decreasing $\sigma$ with increasing $\Lbol$ is changed to increase at around $\EddRatio = 0.1$ due to the shielding effect. Similarly, at a given $\Lbol$, because of the shielding effect, the torus with small $\MBH$ can have a large angular width despite the strong radiation, leading to $\sigma$ that becomes thin and then thick again with $\MBH$.

In Fig.~\ref{fig:frac_Eddington_ratio}, we also examine the fraction of type 1 AGNs as a function of the Eddington ratio based on panel (c) of Fig.~\ref{fig:frac_clumpy}. The top panel demonstrates the type 1 AGN fraction with respect to the Eddington ratio for different ranges of $L_{\rm bol}$, and similarly the bottom panel indicates different ranges of $M_{\rm BH}$ as labeled in different colors. Symbols indicate observational results and dashed lines are taken from the model of panel (c) in Fig.~\ref{fig:frac_clumpy}. We only consider bins that contain at least 30 galaxies. In the top panel, the type 1 AGN fraction varies as a function of the Eddington ratio, showing rise and fall trends with $\lambda_{\rm Edd}$. However, there are offsets between the trends depending on the range of $\Lbol$ and the overall fractions are large for luminous AGNs. In the bottom panel, the trends with $\lambda_{\rm Edd}$ change similarly to those in the top panel, but the overall fractions become large for massive black holes.
 
General trends of the observation seem consistent with the model despite some points deviating from the prediction. Again, the discrepancies between the observations and the model predictions at $\lambda_{\rm Edd} \lesssim 10^{-3}$ can be explained by considering the disappearance of the BLR at the low Eddington ratio \citep[e.g.,][]{nic00,lao03,nic03,eli09}. Equations~(\ref{eqn:sigma}) and (\ref{eqn:N0}) can be re-written in terms of the Eddington ratio: $\sin\sigma = -0.25+0.25\,\sqrt{1+1.3 \times 10^{-2}  \,e^{N_0/e}/\lambda_{\rm Edd}}$ for equation~(\ref{eqn:sigma}) and $N_{\rm 0} = 19.2 \,(\lambda_{\rm Edd})^{0.30} (L_{45})^{-0.08}$ or $15.7 \,(\lambda_{\rm Edd})^{0.22} (M_{8})^{-0.08}$ for equation~(\ref{eqn:N0}). According to the model, as $\lambda_{\rm Edd}$ increases, $N_0$ monotonically increases while $\sigma$ becomes thin and then thick again as shown in Fig.~\ref{fig:model_N0_sigma_frac}. Therefore, with the two quantities competing, the type 1 AGN fraction shows the rise and fall trend. Moreover, at a given $\lambda_{\rm Edd}$, $N_0$ decreases (and thus $\sigma$ decreases) as $\Lbol$ or $\MBH$ increases. This leads to large fractions for high luminosity or massive black holes as shown in Fig.~\ref{fig:frac_Eddington_ratio}.

Recent studies show that the local dynamically measured black hole sample may be biased, raising the normalization of the scaling relations derived from the sample by a factor of a few \citep[e.g.,][]{rei15,las16,sha16}. Therefore, we test the impact of the possible bias by lowering the normalization of the $M_{\rm BH}-\sigma_*$ and $M_{\rm BH}-L_{\rm bulge}$ relations, and $f$ in equation~(\ref{eqn:M_BH_Halpha}) by a factor of 3, 5 and 10, and compare results with those in Fig.~\ref{fig:frac_clumpy}. We find the trends of the fractions for different normalization are similar to those shown in Fig.~\ref{fig:frac_clumpy}, except for trends shifted toward smaller $\MBH$. The smaller $\MBH$ leads to more AGNs with the Eddington ratio greater than 1, and almost half of AGNs seem to have luminosity greater than the Eddington limit when we lower the normalization by a factor of 10. In the case of the first and second columns of Fig.~\ref{fig:frac_clumpy}, the model fits get better (lower chi-square values) for lower normalization, while getting worse in the third column. Though there are some differences in both observations and model fits, the fitting parameters do not change significantly compared to those in Table~\ref{tab:fitting}: $\tilde{N_{\rm 0}} = 9.3 - 10.7$, $a =$ -$0.24$ $-$ -$0.17$ and $b = 0.14 - 0.18$. This means that the interpretation of our model is still valid and meaningful despite the controversial issues on the scaling relations of $\MBH$.

\section{Summary}
\label{sec:summary}
In this study, we explore the structure of the AGN torus through the fraction of optically-selected type 1 AGNs from the SDSS database. Our target galaxies are from the main galaxy sample ($r_{\rm petro} < 17.77$) in the redshift range of $0.01 \leq z \leq 0.2$. For type 1 AGNs, we use the up-to-date type 1 AGN catalog provided by \citet{oh15}. For type 2 AGNs, the selection is performed based on the BPT diagram. In addition to considering Seyfert 2 galaxies, we introduce a demarcation line derived from the relation between the SFR and $\OIII$ luminosity to identify type 2 AGNs from the LINER and the composite region.  As a result, we have 4,268 type 1 AGNs and 15,158 type 2 AGNs, resulting in a ratio of approximately 1:4 between the two types. The primary goal of this study is to determine how different methods of deriving $\MBH$ and $\Lbol$ can affect the type 1 AGN fraction in the $\MBH-\Lbol$ plane, and to interpret the trends using an analytic torus model based on the clumpy-torus scheme. The main results are summarized as follows:

\begin{enumerate}[leftmargin=*] \itemsep.5pt
\item The increase in the type 1 AGN fraction with $\OIII$ luminosity is observed, as in previous studies. However, the fraction in our study is slightly higher than that in \citet{sim05}, but lower than that in \citet{oh15}, because of different sample selection methods. The fit for our sample with the receding torus model is poor, indicating that the luminosity-dependent model is insufficient to explain our observations.

\item Bolometric luminosity, $\Lbol$, is estimated from $\OIII$ luminosity without reddening correction \citep{hec04} and from $\OI$ and $\OIII$ luminosities with reddening correction \citep{net09}. The two different methods agree on their results for $\Lbol$, except for a small discrepancy at low luminosity. When deriving black hole mass $\MBH$, we consider two $M_{\rm BH}-\sigma_*$ relations and two $M_{\rm BH}-L_{\rm bulge}$ relations, which yield different $\MBH$ values. By comparing the $\MBH$ derived from the scaling relations with the single-epoch H$\alpha$-based $\MBH$ estimates for type 1 AGNs, we find that the $M_{\rm BH}-\sigma_*$ relation in \citet{kor13} and the $M_{\rm BH}-L_{\rm bulge}$ relation in \citet{mcl02} present consistent results to some extent. In the case of the $M_{\rm BH}-\sigma_*$ relation, however, there are differences at the high- and low-mass ends which can be ascribed to uncertainties in velocity dispersion.

\item We explore the type 1 AGN fraction in the $\MBH-\Lbol$ plane using $\MBH$ and $\Lbol$ derived from various methods. The general trends of the fraction are similar for all cases, becoming higher for massive black holes and high luminosities. In terms of the different derivation methods for $\Lbol$, the trends of the type 1 AGN fraction are similar regardless of how $\Lbol$ is estimated. The different methods used to compute $\MBH$, however, yield different trends for the fraction. In particular, the ridge-shaped distributions found by \citet{oh15} are reproduced if we derive $\MBH$ using similar methods used by \citet{oh15}, but are not obvious when we use other methods to calculate $\MBH$.

\item To figure out the behaviors of the type 1 AGN fraction, we develop a model based on anisotropic radiation and the clumpy-torus scheme. The  best-fits of the model present the ridge-shaped features for all cases. In the cases where we calculate $\MBH$ using the $M_{\rm BH}-\sigma_*$ or $M_{\rm BH}-L_{\rm bulge}$ relation, there is no clear ridge-shaped distribution of the type 1 AGN fraction in the observation, and thus, the fits are poor. In the meantime, when we use equation~(\ref{eqn:M_BH_Halpha}) for type 1 AGNs and the $M_{\rm BH}-\sigma_*$ relation for type 2 AGNs, the ridge-shaped distributions from the observation and the model agree well, giving good $\chi^2_\nu$.

\item The best-fit model suggests that the number of the clumpy clouds along a radial equatorial ray $N_0$ decreases with $\MBH$ while $N_0$ increases with $\Lbol$. The angular width of the torus $\sigma$ does not change monotonically; it decreases with increasing $\MBH$ or $\Lbol$ and then turns into an increasing trend at around $\EddRatio=0.1$. The changes in $\sigma$ seem to primarily affect the behavior of the type 1 AGN fraction when $\lambda_{\rm Edd}>10^{-3}$. In the regions where $\lambda_{\rm Edd}<10^{-3}$, the increase in the type 1 AGN fraction in the model, which is not consistent with the observation, is thought to have occurred mainly because of the decrease in $N_0$. However, if we take into account the absence of the BLR at low $\lambda_{\rm Edd}$, we can explain the observed type 1 AGN fraction, which is close to $0$ at $\lambda_{\rm Edd}<10^{-3}$.
\end{enumerate}

Many investigations have been performed at various wavelengths on the type 1 AGN fraction and point out the same direction that the luminosity from the central source of the AGN is a key factor in forming the structure of the torus. \citet{oh15} further proposed that $\MBH$ as well as $\Lbol$ plays a separate role in determining the fraction in the $\MBH-\Lbol$ plane. We use a naive approach to develop an analytic model based on the clumpy-torus scheme and successfully reproduce the ridge-shaped distribution of the fraction found in \citet{oh15}. Our approach is probably over-simplistic and the implication is almost trivial; yet, the apparent success still seems encouraging.

\acknowledgments
We are greatly indebted to Kyuseok Oh for providing his new AGN catalog and reading the early drafts.
S.K.Y. acknowledges support from the Korean National Research Foundation (NRF-2017R1A2A1A05001116). This study was performed under the umbrella of the joint collaboration between Yonsei University Observatory and the Korean Astronomy and Space Science Institute. 

\addtocontents{toc}{\vspace{0.8in}}
\addcontentsline{toc}{section}{BIBLIOGRAPHY}
{

\end{document}